\documentclass[manuscript,nonacm]{acmart}
\AtBeginDocument{%
  \providecommand\BibTeX{{%
    \normalfont B\kern-0.5em{\scshape i\kern-0.25em b}\kern-0.8em\TeX}}}

\setcopyright{acmcopyright}
\copyrightyear{2018}
\acmYear{2018}
\acmDOI{10.1145/1122445.1122456}

\acmConference[PODC '20]{PODC '20: ACM Symposium on Principles of Distributed Computing}{August 03--07, 2020}{Salerno, Italy}
\acmBooktitle{PODC '18: ACM Symposium on Principles of Distributed Computing,
  August 03--07, 2020, Salerno, Italy}
\acmPrice{15.00}
\acmISBN{978-1-4503-XXXX-X/18/06}

\usepackage{graphicx}
\usepackage{amsmath}
\usepackage{amsthm}
\usepackage{amsfonts}
\usepackage{amssymb}
\usepackage{color}
\usepackage{subfigure}
\usepackage{afterpage}
\usepackage{microtype}
\usepackage{enumitem}
\usepackage{mathtools}
\usepackage{appendix}
\appendixtitleon

\usepackage[linesnumbered,ruled, vlined]{algorithm2e}
\SetKwProg{code}{Code}{}{}
\SetKwComment{Comment}{$\triangleright$\ }{}

\makeatletter
\newcommand{\nosemic}{\renewcommand{\@endalgocfline}{\relax}}
\newcommand{\dosemic}{\renewcommand{\@endalgocfline}{\algocf@endline}}
\makeatother

\SetKwRepeat{Do}{do}{whileas}
\SetKwFor{Forpar}{for each}{do in parallel}{}
\SetKwFor{Foreach}{for each}{do}{}

\SetKwFor{Upon}{upon}{do}{}

\SetKwFor{Wait}{wait until}{then}{}
\SetKwFor{Waitnodo}{wait until}{}{}

\let\oldnl\nl
\newcommand{\nonl}{\renewcommand{\nl}{\let\nl\oldnl}}

\newtheorem{definition}{Definition}
\newtheorem{definitions}[definition]{Definitions}
\newtheorem{theorem}{Theorem}
\newtheorem{lemma}[theorem]{Lemma}

\newtheoremstyle{claimstyle}
{2pt} 
{2pt} 
{} 
{} 
{\bfseries} 
{:} 
{.5em} 
{} 

\theoremstyle{claimstyle} 
\newtheorem{claim}{Claim}

\newcommand{\fbar}{\bar{F}}

\newcommand{\cala}{\mathcal{A}}

\newcommand{\calm}{\mathcal{M}}
\newcommand{\caln}{\mathcal{N}}

\newcommand{\calp}{\mathcal{P}}

\newcommand{\commentall}[1]{}

\newcommand{\M}{U}
\newcommand{\m}{\mu}

\newcommand{\point}[3]{#1\stackrel{#3}{\longrightarrow}{#2}}
\newcommand{\notpoint}[3]{#1\stackrel{#3}{\not\longrightarrow}{#2}}

\newcommand{\propagate}[3]{#1\stackrel{#3}{\rightsquigarrow}{#2}}

\newenvironment{proofofclaim}{\noindent{\emph{Proof of Claim:}}}{}

\newcommand{\red}[1]{{\color{red}{#1}}}

\begin{document}

\title{Asynchronous  Byzantine Approximate Consensus in 
	Directed Networks}


\author{Dimitris Sakavalas}
\email{dimitris.sakavalas@bc.edu}
\author{Lewis Tseng}
\email{lewis.tseng@bc.edu}
\affiliation{%
  \institution{Boston College}
  \country{USA}
  }

\author{Nitin H. Vaidya}
\affiliation{%
  \institution{Georgetown University}
    \country{USA}}
\email{nitin.vaidya@georgetown.edu}


\begin{abstract}
 In this work, we study the  approximate consensus problem in asynchronous message-passing networks where some nodes may become Byzantine faulty. We answer an \textit{open problem} raised by Tseng and Vaidya, 2012, proposing the \textit{first algorithm of optimal  resilience} for directed networks. Interestingly, our results show that the \textit{tight} condition on the underlying communication networks for  
asynchronous Byzantine approximate consensus coincides with the tight condition for synchronous Byzantine exact consensus. Our results can be viewed as a non-trivial generalization of the algorithm by Abraham et al., 2004, which applies to the special case of complete networks. The tight condition and techniques identified in the paper shed light on the fundamental properties for solving approximate consensus in asynchronous directed networks.
\end{abstract}




\maketitle

\section{Introduction}
\label{sec:intro}

%
%
%

The extensively studied fault-tolerant consensus problem~\cite{lamport_agreement}
is a fundamental building block of many important distributed computing applications \cite{AA_nancy}. The FLP result~\cite{FLP85} states that it is impossible to achieve \emph{exact} consensus in asynchronous networks where nodes may crash (exact consensus requires nonfaulty nodes to reach an agreement on an identical value). The FLP impossibility result led to the study of weaker variations, including \emph{approximate consensus}~\cite{AA_Dolev_1986}. With approximate consensus,
nonfaulty nodes only need to output values that are within $\epsilon$ of each other for a given $\epsilon > 0$. Practical applications of approximate consensus range from  sensor fusion~\cite{BS92} and  load balancing~\cite{Cybenko89}, to  natural systems like flocking~\cite{VCBCS95} and opinion dynamics~\cite{Hegselmann02}.
The feasibility of achieving consensus depends on the type of faults considered in the system. The literature has mainly focused on crash and Byzantine faults, the latter being the worst case since the misbehavior of faults may be arbitrary.  In this work, we focus on the asynchronous Byzantine approximate consensus problem under the existence of at most $f$ faults.
  
Another important parameter affecting the feasibility is the topology of the underlying communication network $G=(V,E)$ in which nodes represent participants that reliably exchange messages through edges. The relation between network topology and feasibility in undirected networks was studied shortly after the introduction of the respective problems (e.g., \cite{AA_nancy,dolev_82_BG}). For $|V|=n$, connectivity $\kappa(G)$ of the network  and upper bound $f$ on the number of faults, Table \ref{t:undirected} summarizes the well-known necessary and sufficient topological conditions for achieving exact consensus and approximate consensus in various settings where $G$ is \textit{undirected}. In undirected networks,
satisfying the necessary graph conditions in Table \ref{t:undirected} also implies
feasibility of reliable message transmission (RMT) (cf.~\cite{DDWY93}), which can be exploited to
simulate algorithms designed for complete networks.
%
%
\begin{table}[h]
\begin{center}
\begin{tabular}{|p{4cm}|p{4cm}|p{4cm}|} \hline
& Crash fault & Byzantine fault \\ \hline
Synchronous system \newline
(exact consensus) & $n>f$ and $\kappa(G)>f$\newline \cite{AA_nancy} & $n>3f$ and $\kappa(G)>2f$\newline \cite{dolev_82_BG}
 \\ \hline
Asynchronous system \newline
(approximate consensus) & $n>2f$ and $\kappa(G)>f$\newline  \cite{AA_nancy} & 
$n>3f$ and $\kappa(G)>2f$  \newline \cite{impossible_proof_lynch, abraham_04_3t+1_async, DDWY93}
\\
 \hline
\end{tabular}
\end{center}
\caption{Necessary and Sufficient Conditions for Undirected Graphs}
\label{t:undirected}
\end{table}

The study of consensus in directed graphs is largely motivated by wireless networks wherein different nodes may have different transmission range, resulting in directed communication links. While the necessary and sufficient conditions for undirected graphs have been known for many years, their generalizations  for directed graphs appeared only after 2012, e.g., \cite{TV12arxiv,Tseng_podc2015,Sundaram_journal,vaidya_PODC12}. This is mainly due to the fact that no direct relation appears between reliable message transmission and consensus in directed graphs. 

As Table~\ref{t:directed} summarizes, for directed graphs, Tseng and Vaidya \cite{Tseng_podc2015,TV12arxiv} obtained necessary and sufficient conditions for solving consensus in the presence of crash faults in synchronous and asynchronous systems both. However, they were able to obtain such conditions for Byzantine faults only for synchronous systems. 
The determination of a tight condition for  the asynchronous Byzantine model remains open since 2012. This paper closes this gap in the results. We identify a family of \textit{new} conditions which we prove equivalent to the ones obtained in \cite{TV12arxiv,Tseng_podc2015}, offering an important intuition, which essentially leads to the answer of this open question. Our condition family consists of  {\em 1-reach}, {\em 2-reach} and {\em 3-reach} conditions, which are later defined in Section~\ref{sec:prelim}.\footnote{The general k-reach condition family,  presented in the appendix, encompasses conditions 1-reach, 2-reach, 3-reach and may be of further interest.}
Results from \cite{TV12arxiv,Tseng_podc2015} imply that the {\em 3-reach} condition is tight for exact Byzantine consensus in synchronous systems.
A key contribution of this paper is to show that the {\em 3-reach} condition is \textit{also} necessary and sufficient for asynchronous Byzantine consensus in directed graphs.
\begin{table}[h]
\begin{center}
\begin{tabular}{|p{4cm}|p{5cm}|p{5cm}|} \hline
& Crash fault & Byzantine fault \\ \hline
Synchronous system \newline
(Exact consensus) & {\em 1-reach} condition (see Section \ref{sec:prelim}) \newline Tseng and Vaidya 2015 \cite{Tseng_podc2015} & {\em 3-reach} condition (see Section \ref{sec:prelim}) \newline Tseng and Vaidya 2015 \cite{Tseng_podc2015} \\ \hline
Asynchronous system \newline
(Approximate consensus) & {\em 2-reach} condition (see Section \ref{sec:prelim}) \newline Tseng and Vaidya 2012, 2015 \cite{TV12arxiv,Tseng_podc2015} & {\bf 3-reach condition (this paper)}\newline \textbf{open  problem since 2012}
\\ \hline
\end{tabular}
\end{center}
\caption{Necessary and Sufficient Conditions for Directed Graphs}
\label{t:directed}
\end{table}

\begin{figure}
	\centering     
	\subfigure[Byzantine exact consensus feasible for $f=1$]{\label{fig:a}\includegraphics[width=49mm]{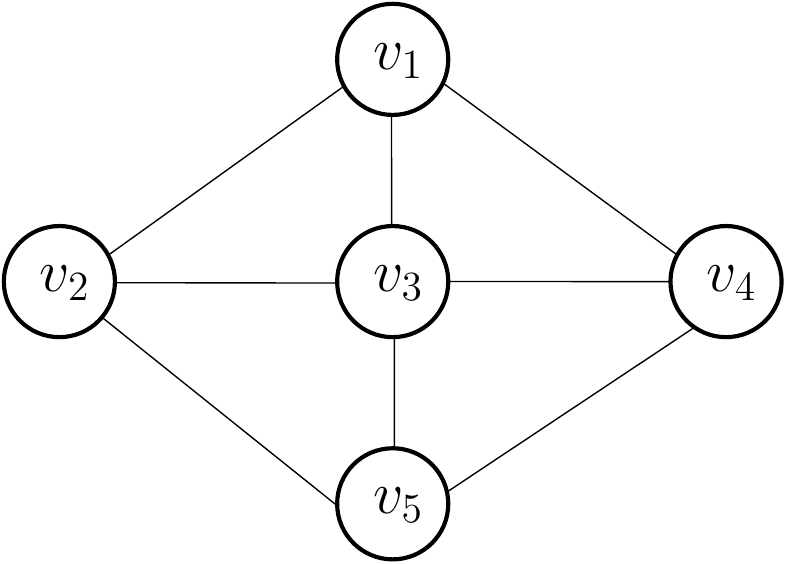}}
	~\hspace{25pt}
	\subfigure[Byzantine exact consensus feasible for $f=2$]{\label{fig:b}\includegraphics[width=9.5cm]{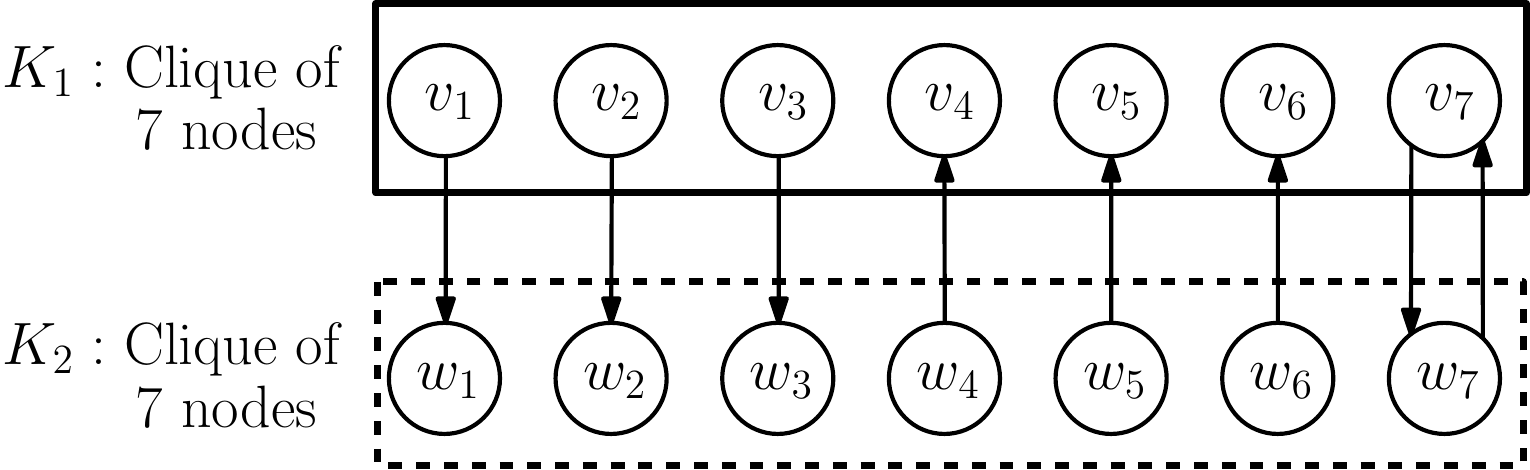}}
	\caption{Example graphs allowing synchronous exact Byzantine  consensus.}
\end{figure}

Essentially, obtaining the tight graph conditions for directed graphs is much more difficult than the undirected case, since consensus may be possible even if reliable message transmission (RMT) is \textit{not} possible between every pair of nodes. This is unlike the case of undirected graphs, as observed previously. For instance, Figure~\ref{fig:a} presents an undirected network, where synchronous exact Byzantine consensus is possible for  $f=1$. In this graph,  all-pair RMT is possible, since $\kappa(G)>2f$ allows any pair of nodes to communicate through at least $2f+1=3$ disjoint paths.
 Note that removing any edge will reduce  $\kappa(G)$, which will make both RMT and consensus impossible. 
%
Such an all-pair RMT is not necessary in directed graphs. In particular, Figure \ref{fig:b} shows a network that satisfies the {\em 3-reach} condition (stated later in Section \ref{sec:prelim}) -- this network includes two cliques, each containing 7 nodes, and eight additional directed edges as shown (edges within each clique are not shown in the figure). Observe that there are pairs of nodes (e.g., $v_1$ and $w_1$) that are connected via only $2f=4$ disjoint paths. Clearly, all-pair RMT is not feasible in this case but consensus can still be achieved, as shown by \cite{TV12arxiv} and our results.  
\commentall{
Observe that there are only two node-disjoint directed paths from node $a$ to node $b$; and similarly, there are only two node-disjoint directed paths from node $b$ to node $a$. Thus, for instance, it is not possible to a priori guarantee reliable communication from node $a$ to $b$ -- whether reliable communication will be feasible depends on which node is faulty. This uncertainty makes the design of algorithms for consensus in directed graphs significantly more complex when compared to undirected graphs.
}
The difficulty posed by directed graphs is further compounded by asynchrony. In this work, we show that the \emph{3-reach} condition is necessary and sufficient  for asynchronous Byzantine approximate consensus in directed graphs -- note that this condition is identical to that proved by Tseng and Vaidya \cite{Tseng_podc2015} for synchronous Byzantine exact consensus.

\paragraph{Related work} Additional related work includes studies of the special class of \emph{iterative algorithms}, which only utilize local knowledge of the network topology and employ local communication between nodes. A tight condition for iterative approximate Byzantine consensus has been presented in~\cite{vaidya_PODC12,Sundaram_journal}. A family of tight conditions for approximate Byzantine consensus under the more general class of  $k$-hop iterative algorithms has been presented recently in~\cite{SV17} but is restricted to synchronous systems. 
The feasibility of asynchronous crash-tolerant consensus with respect to the  $k$-hop iterative  algorithms has been considered in~\cite{STV18_opodis}. A series of works~\cite{LPS13, PPS17, PPS17FCT} studies the effects of topology knowledge on the feasibility of RMT, and consequently exact consensus in undirected networks with Byzantine faults.

%


\section{Preliminaries and Main Result}
\label{sec:prelim}

For the approximate consensus problem~\cite{AA_Dolev_1986}, each node is given a real-valued input, and the algorithm needs to satisfy the three properties below. 

\begin{definition}\label{definition:approx} Approximate consensus is achieved
if the following conditions are satisfied for a given $\epsilon>0$.
	
	\begin{enumerate}
		\item \textit{Convergence}: the output values of any pair of nonfaulty nodes are within $\epsilon$ of each other. 
		
		\item \textit{Validity}: the output of any nonfaulty node is within the range of the inputs of the nonfaulty nodes.
		
		\item \textit{Termination}: all nonfaulty nodes eventually output a value.
	\end{enumerate}
	
\end{definition}

\subsubsection*{System Model}
We consider an asynchronous message-passing network.
The underlying communication network is modeled as a simple {\em directed} graph $G(V, E)$, where $V=\{1,\dots,n\}$ is the set of $n$ nodes, and $E$ is the set of directed edges between the nodes in $V$.
Node $i$ can reliably transmit messages to node $j$ if and only if
the directed edge $(i,j) \in E$.
Each node can send messages to itself as well; however,
for convenience, we {exclude self-loops} from set $E$. A link is assumed to be reliable, but the message delay is not known a priori.

In the system, at most $f$ nodes may become Byzantine faulty during an execution of the algorithm.
A faulty node may {\em misbehave} arbitrarily. The faulty nodes may potentially collaborate with each other.

\subsubsection*{New Graph Conditions}
 Hereafter, we will use the notation $\overline{X}$ to denote the complement $V\setminus X$ of  set $X\subseteq V$.  The subgraph of $G$ induced by node set $Y\subseteq V$ will be denoted by $G_{Y}$. For a given node set $F\subseteq V$,
we now define the \emph{reach set of node $v$ under $F$}, originally introduced in \cite{Tseng_podc2015}. 

\begin{definition}[Reach set of $v$ under $F$]\label{def:reach}
	For node $v\in V$ and  node set $F\subseteq V \setminus\{v\}$, define
	$$reach_v(F)=\{u\in \overline{F} : \text{ $u$ has a directed path to $v$ in graph } G_{\overline{F}}  \}$$
\end{definition}

Observe that a node $u$ belongs to $reach_v(F)$ if $v$ is reachable from $u$  in the subgraph of $G$ induced by node set $V\setminus F$.
Trivially, $v$ is in $reach_v(F)$.
With the definition of a reach set,
we introduce the \mbox{\em 1-reach}, {\em 2-reach} and {\em 3-reach} conditions referred in Section \ref{sec:intro}.
Intuitively speaking,
in the definitions below, the sets $F$, $F_v$, $F_u$ represent {\em potential} sets of faulty nodes; thus, these sets are chosen to be of
size $\leq f$. In the following, recall that $\overline{C}$ denotes the set $V\setminus C$.

\begin{definition}[Reach Conditions]
\label{def:reach1} We define three conditions:
	 \begin{itemize}
		\item \textbf{1-reach:} For any $F\subset V$ such that $|F|\leq f$ and any nodes $u,v\in \overline{F}$, we have $$reach_u(F)\cap reach_v(F)\neq \emptyset.$$
		
		\item \textbf{2-reach:} For any nodes $u, v\in V$ and any node subsets $F_u$, $F_v$ such that $|F_u|, |F_v|\leq f$, $u\in \overline{F_u}$, and $v \in \overline{F_v}$, we have $$reach_v(F_v)\cap reach_u(F_u)\neq \emptyset.$$
		
		\item \textbf{3-reach:}
		For any nodes $u, v\in V$ and any node subsets $F$, $F_u$, $F_v$ such that $|F|, |F_u|, |F_v|\leq f$, $u\in \overline{F\cup F_u}$, and $v \in \overline{F\cup F_v}$, we have
		$$reach_v(F\cup F_v)\cap reach_u(F\cup F_u)\neq \emptyset.$$
	\end{itemize}
	
\end{definition}	 

It is easy to verify that in a clique, 1-reach, 2-reach, and 3-reach are equivalent with $n>f, n>2f$, and $n>3f$ respectively. Details can be found in Appendix~\ref{sec:kreach}.


\subsubsection*{Main Results}
As noted previously, Tseng and Vaidya~\cite{Tseng_podc2015} obtained necessary and sufficient conditions enumerated in Table 2. We have shown, in Appendix~\ref{sec:kreach}, that each of their conditions to be \textit{equivalent} to a respective reach condition in Definition \ref{def:reach1} above. In particular, based on the results in \cite{Tseng_podc2015}, we can prove Theorems \ref{thm:sync:crash}, \ref{thm:async:crash} and \ref{thm:sync:byz} below.
These results are not used to prove our main results; hence, we defer the presentation of the original conditions in \cite{Tseng_podc2015} and the equivalence proofs to Appendix~\ref{sec:kreach}.

\begin{theorem}
\label{thm:sync:crash}
	Synchronous exact consensus is possible in network $G(V,E)$ in the presence of up to $f$ crash faults if and only if $G$ satisfies 1-reach condition.
\end{theorem}

\begin{theorem}
\label{thm:async:crash}
Asynchronous approximate consensus is possible in network $G(V,E)$ in the presence of up to $f$ crash faults if and only if $G$ satisfies 2-reach condition.
\end{theorem}

\begin{theorem}
\label{thm:sync:byz}
Synchronous exact consensus is possible in network $G(V,E)$ in the presence of up to $f$ Byzantine faults if and only if $G$ satisfies 3-reach condition.
\end{theorem}



\noindent
\rule{\textwidth}{2pt}
\subsubsection*{Main Result}
\begin{theorem}
\label{thm:async:byz}
Asynchronous approximate consensus is possible in network $G(V,E)$ in the presence of up to $f$ Byzantine faults if and only if 3-reach is satisfied.
\end{theorem}
This result solves the open problem in Table 2 in Section \ref{sec:intro}.

\noindent
\rule{\textwidth}{2pt}

\noindent{\bf Proving the main result:}
The sufficiency of the {\em 3-reach} condition for asynchronous Byzantine approximate  consensus is demonstrated constructively in Section \ref{sec:algo}, using Algorithm~\ref{alg:BW} for achieving this goal. The necessity of the {\em 3-reach} condition for asynchronous Byzantine approximate  consensus follows by standard indistinguishability arguments; the proof is deferred to Appendix~\ref{sec:necessity}.  

\noindent{\bf Technique Outline:}
 Our result generalizes the result 
 of~\cite{abraham_04_3t+1_async}, which shows the sufficiency of condition $n>3f$ for asynchronous Byzantine  approximate consensus in a clique. Note that the condition coincides with the tight condition for the synchronous case (cf. Table~\ref{t:undirected}). For  directed graphs, we show that 3-reach is the tight condition for both the synchronous and asynchronous cases. 
 Condition {\em 3-reach} states that there exists a node that has (i) a directed path to node $u$ in the subgraph induced by the node subset $\overline{F\cup F_u}$, and also (ii) a directed path to node $v$ in the subgraph induced by the node subset $\overline{F\cup F_v}$. This ``source of common influence'' for any pair of nodes is crucial for achieving consensus. 
 We outline two techniques used towards our generalization, since they may provide useful intuition for other fault-tolerant settings.

\emph{Maximal Consistency:} We simplify the Reliable-Broadcast subroutine of~\cite{abraham_04_3t+1_async} by essentially replacing several rounds of communication between nodes with flooding.
Even in a clique, $r$ communication rounds can be simulated by  flooding through propagation paths~\footnote{Observe that the definition of a path also applies in a clique network.} of length at most $r$. The receiver of all these propagated messages can then detect the existence of faults in certain propagation paths if the propagated values  are inconsistent (i.e., values from different paths do not match).	The technique appears in the use of the Maximal-Consistency condition in Algorithm~\ref{alg:BW}; this simple condition 
provides similar properties as Reliable-Broadcast of~\cite{abraham_04_3t+1_async}. 
 	
\emph{Witness node:} The \emph{witness} technique used in~\cite{abraham_04_3t+1_async} relies on the fact that for any pair of nodes, there is a nonfaulty  witness node which provides them with enough common information. The existence of an analogous nonfaulty witness for directed networks is implied by the 3-reach condition. Intuitively, even if two nodes $v,u$ ``suspect" different sets $F_v, F_u$ to be faulty, the existence of a common nonfaulty witness  guarantees the flow of common information to both.
Guaranteeing that all pairs of nonfaulty nodes gather enough common values while ensuring that nonfaulty nodes with wrongly suspected faulty set are always able to proceed appeared to be the most challenging part of the proposed algorithm. This technique appears in how each node verifies the messages that it has received at line \ref{line:verify_function} in Algorithm \ref{alg:BW}. Generally speaking, a node tries to collect as many ``verified messages'' as possible while it cannot wait for messages that might never arrive (i.e., message tampered by faulty nodes).

\section{Useful Terminology}
\label{sec:useful}

Recall that directed graph $G=(V,E)$ represents the network connecting the $n$ nodes in the system. Thus, $n=|V|$.
We will sometimes use the notation $V(G)$ to represent the set of nodes in graph $G$.
In the following, we will use the terms {\em edge} and {\em link} interchangeably. We now introduce some graph terminology to facilitate the discussion. 

\vspace{3pt}





\begin{itemize}
    \item A path is represented by an ordered list of vertices. In particular, $p=\langle v_1, \ldots, v_k\rangle$ is a directed path $p$ comprising of nodes $v_1, \ldots, v_k\in V$ and directed edges $(v_i, v_{i+1}) \in E$, where $1\leq i\leq k-1$.
    
     \item $\mathbf{init(p)}$ and $\mathbf{ter(p)}$,  will be used to denote the initial node $v_1$ and terminal node $v_k$ of a path $p=\langle v_1, \ldots, v_k\rangle$.

    \item A \textbf{$\mathbf{(v_1, v_k)}$-path} is a path with $init(p)=v_1$ and $ter(p)=v_k$. 
    \item Operation $p||u= \langle v_1, \ldots, v_k, u\rangle$ denotes the concatenation of path $p=\langle v_1, \ldots, v_k\rangle$ with node $u$ assuming that $(v_k,u)\in E$. Analogously,  if $ter(p)=init(p')$, then $p||p'$ denotes the concatenation of paths $p$ and $p'$.   
    
    \item {\bf Redundant path:} a path $p$ is a redundant path if $p=p_1||p_2$ for some simple paths $p_1$ and $p_2$ ($p_1$ and $p_2$ have no cycles) and one of $p_1, p_2$ may be empty. 
    Note that a redundant path may contain cycles and its length is upper bounded by $2n$.
    
    \item The set of all redundant paths in graph $G_{Y}$ (defined above) will be denoted as $\calp^r_{Y}$.
    
    \item {\bf Fully nonfaulty path:} a path consisting entirely of nonfaulty nodes. 
    
    \item {\bf $\mathbf{(A,v)}$-paths:} given a set $A\subseteq V$ and a node $v\in V$, an  $(A,v)$-path $p$ is a path with $init(p)\in A$ and $ter(p)=v$.\medskip
    
    When convenient, we will interpret a path $p$ as the set of nodes in the path. The next few definitions use this interpretation for a node set $C$ and paths $p=\langle v_1, \ldots, v_k\rangle$, $p'=<v'_1, \ldots, v'_k>$.\smallskip
    
    \item  $C\cap p$ will denote the intersection $C\cap\{v_1, \ldots, v_k\}$.
    \item We will say that $p\subseteq C$ if $\{v_1, \ldots, v_k\}\subseteq C$.
    \item By $p\cap p'$, we will denote the node intersection $\{v_1, \ldots, v_k\}\cap \{v'_1, \ldots, v'_k\}$ of paths $p$ and $p'$.
    
	\end{itemize}

\noindent\hrule

\vspace*{4pt}

\begin{definition}[$f$-cover of a path set]
For a set of paths $P$, a node set $C$ is a \emph{$f$-cover} of $P$, if $|C|\leq f$, and \\$\forall p\in P,~~\text{  }~~C\cap p\neq \emptyset$.
\end{definition}

\begin{definition}[Reduced Graph]
	\label{def:reduced}
	For graph $G=(V,E)$, and sets $F_1,\,F_2\subseteq V$, such that $|F_1|,|F_2|\leq f$, reduced
	graph $G_{F_1,F_2}=(V,E_{F_1,F_2})$ has set of vertices $V$, and the set
	of edges $E_{F_1,F_2}$ is obtained by removing from $E$
	all the outgoing links at each node in $F_1\cup F_2$. 
	That is, $$E_{F_1,F_2}~=~E~\setminus~ \left\{(u,v)~|~u\in F_1\cup F_2,~v\in V,~v\neq u\right\}.$$
\end{definition}

\begin{definition}[Source Component]\label{def: sourcecomponent}
	For graph $G=(V,E)$, and sets $F_1,\,F_2\subseteq V$, such that $|F_1|,|F_2|\leq f$, source component $S_{F1,F2}$ is defined as the set of those nodes in the reduced graph $G_{F1,F2}=(V,E_{F_1,F_2})$ that have directed paths to all the nodes in $V$.
\end{definition}

By definition, the nodes in $S_{F1,F2}$ form a strongly connected component in $G_{F1,F2}$. The source component $S_{F1,F2}$ has other desirable properties that will be introduced when we prove the correctness of our algorithm later.

\section{Asynchronous approximate consensus in directed networks}
\label{sec:algo}

We next present an algorithm for approximate Byzantine consensus in asynchronous directed networks. 
The algorithm is optimal in terms of resilience, meaning that it matches the impossibility condition of the problem, i.e., the algorithm works in any graph that satisfies 3-reach.
Our solution is inspired by the asynchronous approximate consensus algorithm of~\cite{abraham_04_3t+1_async} as explained in Section~\ref{sec:prelim}. 
However, the tools used in the  algorithm of~\cite{abraham_04_3t+1_async} prove highly non-trivial to generalize in the case of a partially-connected directed network. This is due to the constraint of directed edges. In complete graphs considered in~\cite{abraham_04_3t+1_async}, information can flow both directions, and each node can use the same rule to collect information. In the case of directed networks, information may only be able to flow in one direction.
We need to devise new tools for nodes to exchange and filter values so that enough common information is shared between any pair of nonfaulty nodes.

\paragraph{Outline of the algorithm}
In our algorithm, each node $v$ maintains a state value $x_v[r]\in \mathbb{R}, \text{ for } r\in \mathbb{N}$, which is updated regularly, with $x_v[0]$ denoting the real-valued input of node $v$. 
Value $x_v[r]$ represents the $r$-th update of the state value of node $v$; we will also refer to it as the state value of $v$ in (asynchronous) \emph{round}  $r$. Observe that in asynchronous systems, $v$ updates the value every time it receives enough messages of a certain type (i.e., an event-driven algorithm),  thus creating the sequence $\left(x_v[r]\right)_{r\in\mathbb{N}}$. The $r$-th value update of a node $v$ may happen at a different real time than the respective update of another node $u$. 
	
The proposed algorithm is structured in two parts presented in Algorithm~\ref{alg:BW}: Byzantine Witness (BW) and Algorithm~\ref{alg:filteraverage}: Filter-and-Average (FA). Algorithm BW intuitively guarantees that all nonfaulty nodes will gather enough common information in any given (asynchronous) round. 
The value update in each round is described in Algorithm FA, where we propose an appropriate way for a node to filter values received in Algorithm BW and obtain the state value for next round as an average of the filtered values.  
Each node needs to filter potentially faulty values to guarantee validity.

\subsection{Algorithm Preliminaries}

The proposed algorithm utilizes the propagation of values through certain \textit{redundant paths} (defined in Section \ref{sec:useful}).
We then describe tools for handling information received by a node through different paths 

\paragraph{Messages and Message Sets.} In our algorithm, the messages propagated are of the form $(x,p)$ where $x$ is the propagated value and $p$ corresponds to the (redundant) path through which the value $x$ is propagated, i.e., its \emph{propagation path}. For a message $m=(x,p)$, we will use the notation  $value(m)=x$ and $path(m)=p$ to denote the propagated value and propagation path, respectively. For simplicity, we will also use the terminology \emph{$v$ receives value $x$ from $u$} whenever node $v$ receives $x$ through some path $p$ initiating at node $u$.  A \textit{message set} $\calm$ is a set of messages of the form $m=(x,p)$ where $x$ is the value reported though propagation path $p$. Given $\calm$, we will use $\calp(\calm)$ to denote the set of \textit{all propagation paths} in $\calm$, i.e., 
$$\calp(\calm)=\{p~:~ (x,p)\in \calm\}$$

As defined below, given a node set $A$ and a message set $\calm$, the \emph{exclusion} of $\calm$ on $A$  consists of the messages of $\calm$
that are propagated on paths that \underline{do not include any
node in  $A$.}

\begin{definition}[Exclusion of message set]	Given a message set $\calm$ and $A\subseteq V$, the exclusion of $\calm$ on $A$ is the  set
$$\calm|_A=\{(x,p)\in\calm~:~p\cap A=\emptyset\}$$
\end{definition}

 The notions of a \emph{consistent message set} and \emph{full message set}, presented below, are used to facilitate  fault detection. A message set $\calm$ is consistent if all value-path tuples initiating at the same initial node report the same value.

\begin{definition} [Consistent message set]
	A message set $\calm$ is \emph{consistent} if 
	$$\text{for any two value-path pairs}~ (x,p),(x',p')\in \calm,~~~~ init(p)=init(p') ~\Rightarrow~ x=x'$$
\end{definition}

\noindent Given a consistent message set $\calm$, if $(x,p)\in \calm$ and $init(p)=v$, then we
define $value_v(\calm)=x$. That is,
for a node $v$ that appears as an initial node of a path in $\calp(\calm)$,
$value_v(\calm)$ denotes \underline{the \textit{unique} value corresponding to $v$.} Note that the value is guaranteed to be unique owing to the the definition of a consistent message set.

\vspace{5pt}

We say that the received message set $\calm$ is a \emph{full message set} for $(A,v)$, whenever a node $v$ receives messages from \underline{all possible incoming redundant paths excluding node set $A$}. The formal definition follows.
	
	\begin{definition}[Full message set]
		Given $A\subseteq V$ and $v \in V\setminus A$,
		a message set $\calm$ is \emph{full for $(A,v)$} if
		$$\{p\in \calp^r_{\overline{A}} : ter(p)=v\} \subseteq \calp(\calm)$$
	\end{definition}

\subsection{Algorithm \emph{Byzantine Witness} (BW)}
	
The \emph{Byzantine  Witness} (BW) algorithm, presented as Algorithm~\ref{alg:BW}, intuitively guarantees that all nonfaulty nodes will receive enough common state values in a specific asynchronous round of the algorithm; eventually, this common information  guarantees convergence of local state values at all nonfaulty nodes. 
The algorithm is event-driven. That is, whenever a new message is received, each node checks whether a certain set of conditions are satisfied, and whether it should take an action. (In particular, Line 6, Line 8, Line 10, and Line 12 of Algorithm \ref{alg:BW} will be triggered upon receipt of a new message.)

\paragraph{Parallel executions.} For the sake of succinctness, we present a parallel version of Algorithm BW; the algorithm makes use of parallel executions (threads) for any potential fault set $F$. Note that there are exponential number of threads. In the parallel thread for set $F$, a node ``guesses'' that the actual fault set of this execution is $F$, and checks for inconsistencies to reject this guess.   Observe that in lines~\ref{line:ifnextround}-\ref{line:nextroundtrue} of Algorithm BW, the usage of a shared boolean variable $nextround$ guarantees that a node will proceed to the next round during at most one parallel execution; we will later prove that there always exists such a parallel execution that proceeds to the next round. For each round $r$, during this unique parallel execution, the node will execute Algorithm  \emph{Filter-and-Average} (FA), presented as  Algorithm~\ref{alg:filteraverage}, through which the value is updated.   
	
Suppose that a node's parallel thread for set $F'$ proceeds to the next round. It is possible that $F'$ is \textit{not} the actual faulty set.
Our algorithm is designed in a way that even if the guess is incorrect, the node is still able to collect enough common values and make progress. Moreover, the parallel thread for set $F$, where $F$ is the actual fault set, is guaranteed to make progress at every nonfaulty node.
	 
\paragraph{Atomicity} Algorithm BW uses the shared variables $\calm_v$ and $nextround$; $\calm_v$ includes all values received by node $v$ and is updated whenever $v$ receives a new flooded value while $nextround$ indicates if a parallel thread has proceeded to the next (asynchronous) round. For certain parts of the algorithm we need access to shared variables to be atomic, i.e., reads and writes to shared variables can  be performed only by a parallel thread at a time. For clarity of the latter, we make use of the functions \texttt{lock()} and  \texttt{unclock()} which indicate the parts of the code performed in an atomic fashion.

%
%
%
We next describe a flooding subroutine used to propagate state values throughout \textit{redundant paths} in the network.  
	
\paragraph{  \texttt{RedundantFlood} (Redundant Flood) algorithm} In the beginning of each asynchronous round, in algorithm~\ref{alg:BW}, all nodes will flood their values throughout the network. The difference between \texttt{RedundantFlood} algorithm and the standard flooding is that each flooded message is propagated through \underline{any redundant path} in the network, not just through simple paths. The details of the algorithm are deferred to the Appendix~\ref{sec:rfloodalg}. 

	%
	%
	%
	
\paragraph{FIFO flooding of messages} During the execution of the algorithm, we will employ the \texttt{FIFO Flood} and \texttt{FIFO Receive}  procedures which ensure that the order of messages sent from a sender is preserved during reception of these messages by any receiver, when the propagating path is fully comprised of nonfaulty nodes. For the correctness of our algorithm, we only need to FIFO-flood messages through \textit{simple paths} in the network.\footnote{It is possible to use \texttt{RedundantFlood} everywhere. For efficiency, we only use \texttt{RedundantFlood} to propagate state values at the very beginning of each round.} Thus, a node will propagate a message during FIFO-Flood, only if the resulting propagation path does \textit{not} contain any cycle.  We present a high-level description of the \texttt{FIFO Flood} and \texttt{FIFO Receive}  procedures in the Appendix~\ref{sec:fifoflood}.
	
\paragraph{Algorithm BW: Pseudo-code} Algorithm BW is presented in Algorithm ~\ref{alg:BW}. We stress that Algorithm BW is executed for each asynchronous round $r$. Thus, all messages sent in  round $r$ will be tagged with corresponding round identifier $r$. For simplicity, we omit round numbers in the presentation  and the analysis of the algorithm. The properties of Algorithm BW are proved hold for any specific asynchronous round $r$. 
 For brevity, the pseudo-code does not include the termination condition. We defer the discussion on termination to Section \ref{s:correctness_proof}.

	\begin{algorithm}
		\caption{\texttt{BW} (for node $v$ and round $r$) }\label{alg:BW}\medskip
		
		\SetKwFunction{FMain}{AV}
		\SetKwProg{Fn}{Function}{:}{}

		\nonl\hspace{-15pt}\textbf{Input:} State value $x_v$ of node $v$ for round $r$\\
		\Comment*[f]{ Round id $r$, included in all sent messages, is omitted for simplicity}\medskip
		
		\nonl\hspace{-15pt}\textbf{ Code for $v \in V$:}\medskip
		
		\nonl \textbf{Initialization}
		
		$\calm_v\leftarrow\emptyset$
		 \Comment*[f]{ shared variable accessed by all parallel threads}
		 
		 {$nextround\leftarrow false$}
		 \Comment*[f]{ shared variable accessed by all parallel threads}
		 
		 $FIFORec(F)=false$, for each $F\subseteq V$ with $|F|\le f$
		 

		 \bigskip

		\texttt{RedundantFlood} value $x_v$
		
		
		\Forpar({\Comment*[f]{only one parallel thread for some $F_v$ }\\
		\Comment*[f]{executes Filter-and-Average due to lines~\ref{line:ifnextround}-\ref{line:nextroundtrue}}
		}){$F_v\subseteq V\setminus\{v\}$, \text{ with } $|F_v|\le f$\label{line:parallel loop}}{
			
			\Upon{receipt of message $m=(x,p)$}{
				\texttt{lock()}
				
				$\calm_v\leftarrow\calm_v \cup m$ \Comment*[f]{Atomic updates of $\calm_v$}
				
			\texttt{unlock()}
			}\medskip

			\Comment*[f]{Maximal-Consistency Condition
			}\smallskip
			
			\Upon{ ($\calm_v|_{F_v}$ is consistent~~and~~full for $(F_v, v)$ for the first time)\label{line:MConsist}}{
				\texttt{FIFO-Flood} $(\calm_v|_{F_v}, COMPLETE(F_v)) $}\medskip

			\Comment*[f]{FIFO-Receive-All  Condition  for $F_v$ }\smallskip
			
			\Upon{\label{line:FRcondition}{\LARGE(}
				For all $c\in reach_v(F_v)$,~~$v$ \emph{\texttt{FIFO-Receives}} the same message $(\calm^c,COMPLETE(F_v))$ from all simple $(c,v)$-paths $p\subseteq reach_v(F_v)${\LARGE)} \label{line:FIFOreceiveall}
			}{
			
			$FIFORec(F_v)\leftarrow true$
			}
			
				\Upon(\Comment*[f]{For verification of a received value, $v$ will wait  }\\
				\Comment*[f]{to receive the same value from enough paths as implied by Algorithm~\ref{AVfunction}.}){ $Verify(\calm_v, FIFORec(F_v))$\label{line:verifycall}}
				{
					\texttt{lock()}

		\If{$nextround$=false \label{line:ifnextround} }	{Execute Algorithm \texttt{Filter-and-Average}$(\calm_v)$\label{line:executeFA} \Comment*[f]{Execution of Algorithm~\ref{alg:filteraverage}}
				
				$nextround\leftarrow true$ \label{line:nextroundtrue}}
			
			\texttt{unlock()}

						}

		}

\bigskip	

\Fn
{$Verify(\calm_v, FIFORec(F_v))$\label{line:verify_function}}{
	
		$validity\leftarrow false$
		
		\If{$FIFORec(F_v)=true$}{
			
			$validity\leftarrow true$
			
			\Foreach{$(\calm^c, COMPLETE(F_u))$ FIFO-received through a simple $(c,v)$-path $p\subseteq reach_v(F_v)$, with consistent $\calm^c$}{
				$validity\leftarrow validity\wedge Completeness(\calm_v, \calm^c, Fu)$ \Comment*[f]{Completeness: Algorithm~\ref{AVfunction} }

			}

		}
		
		\KwRet $validity$	
	
}

	

%
%
%
%
%
%
%
%
%
%
%
%
%
%

\end{algorithm}	


\begin{algorithm}
	\caption{Function $Completeness(\calm_v,\calm^c, F_u)$} \label{AVfunction}\medskip
	
	\nonl\textbf{Input:} Message sets $\calm_v,\calm^c$, $F_u\subseteq V$\medskip\\
	
		\nonl \textbf{Initialization}
		
		$output\leftarrow true$\medskip

		\Foreach{$F_w\subseteq V$ with $F_w\neq F_u, $ and $|F_w|\le f$}{
		
		\Foreach{ $q\in S_{F_u, F_w}$}{
			$\calm'\leftarrow\{(value_q(\calm^c),p)\in \calm_v : init(p)=q \}$
			\Comment*[f]{All received messages from $q$}\\
			 	\Comment*[f]{ which are consistent with $\calm^c$}
			
			$output\leftarrow output\wedge (\nexists \text{  an $f$-cover } H\subseteq V\setminus S_{F_u,F_w} \text{ of } \calp(\calm'))$

	}
		}

	\KwRet $output$

\end{algorithm}	

%
%
%
%
%
%
%
%
%
%
%
%
%
%
%
%
%
%
%
%

\paragraph{Function $Completeness(\calm_v,\calm^c, F_u)$} 
We first remind the reader that $S_{F_1,F_1}$ denotes the source component of reduced graph $G_{F_1,F_2}$ as defined in Definitions~\ref{def:reduced},~\ref{def: sourcecomponent}.
Observe that due to function $Verify(\calm_v)$ called in line~\ref{line:verifycall}, a node essentially waits to receive additional messages to the ones it received upon considering possible faulty set $F_v$ (during the parallel execution for $F_v$) before it proceeds to update its value through Algorithm \texttt{Filter-and-Average}. Intuitively, for some received message $(M^c, COMPLETE(F_u))$, $v$ waits for the confirmation of the values in $\calm^c$ through enough \textit{redundant paths} from a source component. We will later prove that if message  $(M^c, COMPLETE(F_u))$ is not faulty (i.e., tampered) then $v$ will eventually be able to ``confirm'' the values in $M^c$. For the sake of simplicity, whenever the function $Completeness(\calm_v,\calm^c, F_u)$ at node $v$ is true for some given  $\calm^c, F_u$, we will simply state that condition $Completeness(\calm_v,\calm^c, F_u)$ is satisfied.


\subsection{Properties of Algorithm BW}

In the following, we introduce some notions necessary for the analysis of Algorithm BW. We first borrow the notion of propagation from \cite{TV12arxiv,Tseng_podc2015}.

\begin{definition}[Propagation between sets]\label{def:propagation}
	Given sets $A, B, C\subseteq V$ with $A\cap B=\emptyset$, $B\subseteq C$, set $A$ is said to \emph{propagate in} $C$ to set $B$ if either (i) $B = \emptyset$, or (ii) for each node $b \in B$, there exist at least $f +1$ node-disjoint $(A, b)$-paths
	in the node subgraph of $G$ induced by node set $C$, i.e., $G_C$.
	We will denote the fact that set $A$ propagates in $C$ to set $B$ by  $\propagate{A}{B}{C}$.
\end{definition}

Note that the $f+1$ disjoint paths implied in Definition~\ref{def:propagation}, are entirely contained in $C$.
Next, observe that by Definition \ref{def: sourcecomponent}, for any sets $F_1,F_2\subseteq V$ it holds that $S_{F_1,F_2} = S_{F_2,F_1}$. The following theorem is repeatedly used in our analysis; its proof is based on Corollary 2 in~\cite{TV12arxiv} and the equivalence of 3-reach condition with the condition in \cite{TV12arxiv} (proof in Appendix \ref{sec:kreach}). Intuitively, Theorem~\ref{thm:propagate} below states that if 3-reach is satisfied then there are at least $f+1$ disjoint paths, excluding nodes in $F_1$, that  connect a source component $S_{F_1, F_2}$ with any node outside the source component.

\begin{theorem}\label{thm:propagate}
	Suppose that graph $G=(V, E)$ satisfies condition 3-reach. Then, for any $F_1 \subseteq V$ and $F_2 \subseteq \overline{ F_1}$, such that $|F_1|,|F_2| \le f$, $\propagate{S_{F_1,F_2}}{\overline{ F_1 \cup S_{F_1,F_2}}}{\overline{F_1}}$ and $\propagate{S_{F_1,F_2}}{\overline{ F_2 \cup S_{F_1,F_2}}}{\overline{F_2}}$ hold.
\end{theorem}

Using the notions above, we will next show some important properties of Algorithm BW. As defined in line~\ref{line:MConsist}, we will say that node $v$ satisfies the  \texttt{Maximal-Consistency} Condition for node set $F'$ if it receives the message set $\calm_v$ and $\calm_v|_{F'}$ is consistent and full for $(F', v)$.

\begin{lemma}\label{lemma: MCtermination}
	For any nonfaulty node $v$, the \texttt{Maximal-Consistency} condition will eventually be satisfied during a parallel execution for some set $F'$.
	
\end{lemma}	

\begin{proof}
	Consider $v$'s parallel execution for set $F'=F$, where $F$ is the actual faulty set of the execution.  If the \texttt{Maximal-Consistency} condition has not been satisfied already, it will eventually be satisfied during the parallel execution for $F'=F$. This will happen since every node in $G_{\overline{F}}$ behaves correctly and thus $v$ will eventually receive consistent values from all incoming paths in $G_{\overline{F}}$, i.e., $\calm_v|_{F_v}$ will be consistent and full for $(F_v,v)$. 
\end{proof}

\begin{lemma}\label{lemma:MCconsisteny}
    Consider two nonfaulty nodes $v, u$ that satisfy the \texttt{Maximal-Consistency}  condition on the same set $F'$.
    Let the message sets $\calm_v|_{F'}$ and $\calm_u|_{F'}$ be the sets that are used to pass \texttt{Maximal-Consistency}  condition at $v$ and $u$, respectively.
    Then, the two sets contain the same unique value $\displaystyle x_w, \forall w\in \bigcup_{\mathclap{\substack{F''\neq F'\\
    			F''\subseteq V, |F''|\le f}}}S_{F',F''}$.

	
\end{lemma}

\begin{proof}
    We first prove that for each $w \in  S_{F',F''}$, both nodes $v$ and $u$ will receive a unique value corresponding to $w$, contained in the respective sets $\calm_v|_{F'}$ and $\calm_u|_{F'}$.
	Observe that for any $F''\neq F'$ and $w\in  S_{F',F''}$,  by Theorem~\ref{thm:propagate} and the fact that any source component $S_{F',F''}$ is strongly connected, there exists a simple $(w,v)$-path in $G_{\overline{F'}}$. 
	Since $\calm_v|_{F'}$ is full for $(F',v)$, $\calm_v|_{F'}$ will contain some value $x_w$ corresponding to $w$. Note that this value might \textit{not} be the value sent by $w$, since the above simple path might contain some faulty node.
	Next, recall that we also require $\calm_v|_{F'}$ to be consistent. Therefore, the previously mentioned $x_w$ value contained in $\calm_v|_{F'}$ must be unique . The same argument applies to $\calm_u|_{F'}$, too. 
	
	The  3-reach condition implies the existence of a node $q\in reach_v(F\cup F')\cap reach_u(F\cup F')$ for the actual faulty set $F$. By definition, $q$ is nonfaulty and is connected to both $v,u$ through fully nonfaulty simple paths $p_{q,v}$ and $p_{q,u}$ respectively. 
	By Theorem~\ref{thm:propagate}, either $q\in S_{F',F''}$ or there exist $f+1$ simple disjoint $(S_{F',F''},q)$-paths in $G_{\overline{F'}}$. In both cases, there exists a simple $(w,q)$-path $p_{w,q}$ in $G_{\overline{F'}}$. 
	Note that there might be some faulty nodes in $p_{w,q}$, since $F'$ might \textit{not} be the actual faulty node set.
	
	This observation implies that in $G_{\overline{F'}}$, there exist  a redundant $(w,v)$-path $p_{w,v}=p_{w,q}|| p_{q,v}$ and a redundant $(w,u)$-path $p_{w,u}=p_{w,q}|| p_{q,u}$ such that the first part $p_{w,q}$ is identical in both paths. 
	Note that the $3$-reach condition only implies that $p_{q,v}$ and $p_{q,u}$ are fully nonfaulty. Hence, it is possible that the value sent by node $w$ is $x'_w$, but the message(s) propagated through $p_{w,v}$ and $p_{w,u}$ are different. Since $\calm_v|_{F'}$ and $\calm_u|_{F'}$ are full;  nodes $v$ and $u$ will receive some value from paths $p_{w,v}$ and $p_{w,u}$, respectively. The value received by $v$ and $u$ must be identical. This is because (i) the two redundant paths have a common first part $p_{w,q}$; and (ii) $p_{q,v}$ and $p_{q,u}$ are fully nonfaulty by assumption. Let this value be $x_w$ (which may or may not equal to $x'_w$, the original value sent by $w$). Finally, since $\calm_v|_{F'}$ is \textit{consistent}, all the other messages propagated through paths $p$ with $init(p) = w$ and $ter(p) = v$ must also be $x_w$, the value forwarded by $q$. The same argument applies to $\calm_v|_{F'}$. Thus, For each $w$, there exists a common value $x_w$ in both $\calm_v|_{F'}$ and $\calm_u|_{F'}$.

\end{proof}

Next we prove the main Lemma for the  correctness of Algorithm BW.
With \texttt{FIFO-Receive-all} condition we refer to the condition stated in the event handler of line~\ref{line:FIFOreceiveall}.

\begin{lemma}\label{lemma: RCmain}
	Consider a nonfaulty node $v$ such that the \texttt{FIFO-Receive-All} condition is satisfied at node $v$ for some parallel execution.
	If by the time the \texttt{FIFO-Receive-All} condition is satisfied, $v$ receives $(\calm^c, COMPLETE(F_u))$ from a fully nonfaulty path $p$ with $init(p)=c$, then $v$ will eventually receive a message set $M_v$ such that the
	 $Completeness(\calm_v, \calm^c, F_u)$ condition will  be satisfied at node $v$. 
\end{lemma}

\begin{proof}First observe that since path $p$ is fully nonfaulty,  $c$ is nonfaulty; also lines~\ref{line:parallel loop},\ref{line:MConsist} of BW imply that $c\notin F_u$ since it propagates $(\calm^c, COMPLETE(F_u))$.
	Consider any $F_w\neq F_u$ with $|F_w|\le f$ and any $q\in S_{F_u,F_w}$. Let $F$ be the actual faulty set of the execution. Note that since nonfaulty $v$ receives $(\calm^c, COMPLETE(F_u))$ from a fully nonfaulty path with $init(p)=c$, node $c$ must have FIFO-Flooded this message during the execution. By the 3-reach condition of Definition~\ref{def:reach}, we have the following. 
	\begin{equation}
	\label{eq:3reachalt}
	\forall F'\subseteq V\setminus S_{F_u,F_w}\setminus \{v\}, \text{ such that } |F'|\le f, \exists z \in reach_v(F\cup F')\cap reach_c(F\cup F_u)
	\end{equation}


	This, for any $F'\subseteq V\setminus S_{F_u,F_w}$, implies the existence of a fully nonfaulty simple $(z,v)$-path $p_{z,v}$ and a fully nonfaulty simple $(z,c)$-path $p_{z,c}$ in graphs $G_{\overline{F\cup F'}}$ and $G_{\overline{F\cup F_u}}$, respectively.\footnote{Set $F'$ is not to be confused with the set $F_v$ during the parallel execution of which $v$ satisfies the \texttt{FIFO-Receive-All} condition. Set $F_v$ is arbitrary in the proof. Note that $v$ might even receive values from paths intersecting with $F_v$ in order to satisfy the $Completeness(\calm_v, \calm^c, F_u)$ condition during its parallel execution for $F_v$. This might occur if $F_v$ is a wrong ``guess'' of the actual fault set.}     We consider the following two cases for $z$,
	
	\begin{itemize}
		\item Case I:  $z\in S_{F_u, F_w}$. 
		
		We first prove the following key claim.
		
		\begin{claim}
		\label{claim:z1}
		Both $v$ and $c$ receive an identical value $x_q$ from node $q$.
		\end{claim}
		
		\begin{proofofclaim}
		    Since $S_{F_u, F_w}$ is strongly connected, there exists a simple $(q,z)$-path $p_{q,z}$ in graph $G_{S_{F_u, F_w}}$. This
		    implies the existence of the redundant $(q,c)$-path $p_{q,c}=p_{q,z}||p_{z,c}$ in $G_{\overline{F_u}}$. Recall that by assumption, $c$ is nonfaulty and FIFO-Floods $COMPLETE(F_u)$. This means that $c$ has received a unique value $x_q$ through all redundant $(q,c)$-paths in $G_{\overline{F_u}}$, particularly through path $p_{q,c}$. Note that since $p_{q,z}$ might contain some faulty nodes, $x_q$ might be different from the value originally sent by node $q$.
		    Observe that there also exists the redundant $(q,v)$-path $p_{q,v}=p_{q,z}||p_{z,v}$ in $G_{\overline{F'}}$ which will eventually propagate the same value $x_q$ to $v$. This is because $p_{z,v}$ is fully nonfaulty and the initial part $p_{q,z}$ is identical in both  $p_{q,c}$ and $p_{q,v}$.
		\end{proofofclaim}
		
		\begin{claim}
		\label{claim:z1-f-cover}
		Node $v$ will eventually receive $x_q$ from a set of paths $P$ with no $f$-cover $H\subseteq V\setminus S_{F_u,F_w}$.
		\end{claim}
		
		\begin{proofofclaim}
		    Recall that Claim \ref{claim:z1} holds for any $F'\subseteq V\setminus S_{F_u,F_w}\setminus \{v\}$, with $|F'|\le f$.
		    Then $v$ will eventually receive $x_q$ from all redundant paths $p_{q,z}||p'_{z,v}$ for any $p'_{z,v}$ being a $(z,v)$-path with $p'_{z,v}\cap F=\emptyset$. This is  because all these  $p'_{z,v}$ paths are fully nonfaulty and the initial part $p_{q,z}$ propagates $x_q$ as implied by Claim~\ref{claim:z1}. The set $P$ of all these $p'_{z,v}$ paths does \textit{not} have an $f$-cover $F'\subseteq V\setminus S_{F_u,F_w} $. If there was such an $f$-cover $F'$, this would contradict Equation~(\ref{eq:3reachalt}) because it would mean that no fully nonfaulty $(z,v)$-path would exist in $G_{\overline{F'}}$.\footnote{Observe that if $v\in S_{F_u,F_w}$ and receives $x_q$ from a single path that entirely consists of nodes in $S_{F_u,F_w}$, then no $f$-cover $H\subseteq V\setminus S_{F_u,F_w}$ exists for this path. This is because $H \cap S_{F_u,F_w} = \emptyset$ by definition.}

		\end{proofofclaim}
		
		\item Case II: $z\notin S_{F_u,F_w}$.
		
		Theorem~\ref{thm:propagate} implies that there exist $f+1$ simple disjoint $(S_{F_u,F_w},z)$-paths in $G_{\overline{F_u}}$. This together with the observation that $S_{F_u,F_w}$ is strongly connected imply the existence of $f+1$ simple $(q,z)$-paths $p_1,\ldots,p_{f+1}$ which trivially do not have an $f$-cover $H\subseteq V\setminus S_{F_u,F_w}$.  
		Similarly with the previous case, since $c$ FIFO-Floods $COMPLETE(F_u)$, it must have received the same value from all redundant paths $p_1||p_{z,c},\ldots,p_{f+1}||p_{z,c}$. Since $|F'|\le f$ and $p_{z,v}$ is fully nonfaulty, one of the paths $p_1||p_{z,v},\ldots,p_{f+1}||p_{z,v}$ will also eventually propagate value $x_q$ to $v$. 
		
		Using the same argument for Claim \ref{claim:z1-f-cover} in Case I, $v$ will eventually receive $x_q$ from   a set of paths $P$ with no $f$-cover $H\subseteq V\setminus S_{F_u,F_w}$. 
	\end{itemize}
	
	In both cases, any such value $x_q$ received by $v$ will be consistent with values $\calm^c$ propagated by $c$, and thus $v$ will eventually satisfy the $Completeness(\calm_v, \calm^c, F_u)$ condition.

\end{proof}

In the following, we will consider the case where a node $v$ executes Algorithm Filter-and-Average through line~\ref{line:executeFA}, during its parallel execution for set $F_v$. In this case, $v$ has already satisfied the \texttt{Maximal-Consistency}  condition corresponding to $F_v$ as well as the $Completeness(\calm_v,\calm^c, F_u)$  conditions for all  $COMPLETE(F_u)$ messages it has received by the time it satisfied the \texttt{FIFO-Receive-All} condition. Intuitively, this means that $v$  has received redundant messages corresponding to ``suspicious sets" $F_v,F_u$ by the time it executes Filter-and-Average. Note that there exists only one such parallel execution during which Filter-and-Average is executed; this follows easily from the atomic updates of shared variable $\calm_v$  and lines~\ref{line:ifnextround}-\ref{line:nextroundtrue} of Algorithm BW. 
We will use the following notion in our proofs.

\begin{definition}[Informed node]\label{definition:informed}
	A node $v$ that executes Filter-and-Average during its parallel execution for a set $F_v$ is \emph{informed about set $F_t$} if $F_v=F_t$, or $v$ has satisfied the $Completeness(\calm_v,\calm^c, F_t)$  condition after receiving message $(\calm^c, COMPLETE(F_t))$ from a fully nonfaulty path $p$ with $init(p)=c$. 
\end{definition}

\begin{theorem}\label{thm: FRtermination}
	Any nonfaulty node $v$ will eventually execute Filter-and-Average during a parallel execution for a set $F'$.
	
\end{theorem}

The proof of Theorem~\ref{thm: FRtermination} relies on the observation that algorithm Filter-and-Average will be executed during parallel execution for actual fault set $F$ if not during any other parallel execution. The full proof is presented in Appendix~\ref{sec:BWtermination}.

\begin{theorem}\label{thm:valuesubset}
	
	Let any  pair of nonfaulty nodes $v,u$ which  execute Filter-and-Average during their parallel executions for sets $F_v$ and $F_u$, respectively. Then, both nodes $v$ and $u$ will be informed about a node set $F_t\in \{F_v,F_u\}$, where $t\in \{v,u\}$,  and  will both  receive a common value $ x_q$ for each $\displaystyle q\in \bigcup_{\mathclap{\substack{F_w\neq F_t\\
				|F_w|\le f}}}S_{F_t,F_w}$. More specifically, each value $x_q$ will be the unique value corresponding to node $q$ that node $t$ received by the time it satisfied its \texttt{Maximal-Consistency}  condition.

\end{theorem}

\begin{proof}

	Theorem~\ref{thm: FRtermination} implies that sets $F_v, F_u$ are well defined. Observe that, if $F_v=F_u$ the theorem holds trivially due to Definition~\ref{definition:informed} and Lemma~\ref{lemma:MCconsisteny}; this is because both nodes will trivially be informed about the same set, according to the first part of the definition of an informed node.  Thus, we focus on the case where $F_v\neq F_u$.  The \texttt{FIFO-Receive-All} condition for nodes $v$ and $u$ is satisfied in the parallel execution for $F_v,F_u$ respectively by assumption.  Let $F$ be the actual fault set. Due to the 3-reach condition, there exists a node $c\in reach_v(F\cup F_v)	\cap reach_u(F\cup F_u)$ which is nonfaulty by definition of the \emph{reach} set and is connected to $v,u$ through fully nonfaulty simple paths $p_{c,v}$ and $p_{c,u}$ respectively. Note that both nodes $v,u$ will only satisfy their \texttt{FIFO-Receive-All} conditions only if they receive messages of the form $(\calm^c_v,COMPLETE(F_v))$ and $(\calm^c_u, COMPLETE(F_u))$, respectively,  from $c$ through the existing nonfaulty paths $p_{c,v}, p_{c,u}$ respectively; this holds since due to Line~\ref{line:FRcondition}, node $v$ (analogously node $u$) will wait until it receives $(\calm^c_v,COMPLETE(F_v))$ from all paths entirely comprising of nodes in $reach_v(F_v)$ which include all nodes on $p_{c,v}$. Thus, $c$ must have sent both $COMPLETE(F_v)$, $COMPLETE(F_u)$ messages.  Since these messages are FIFO-flooded from $c$ and there are fully nonfaulty paths connecting $c$ with both $u,v$, one of the two nodes will receive both $COMPLETE(F_v)$, $COMPLETE(F_u)$ messages before satisfying the \texttt{FIFO-Receive-All} condition. Assume without loss of generality that this node is $v$. We will then show that the theorem holds for  $F_t=F_u$.

Similar arguments~\footnote{This follows from the first paragraph of the proof  of Lemma~\ref{lemma:MCconsisteny} for $c$ being any of the nodes $v,u$.} to the ones used in the proof of Lemma~\ref{lemma:MCconsisteny} imply that $c$ must have received a unique value $x_q$ for each $\displaystyle q\in \bigcup_{\mathclap{\substack{F_w\neq F_u\\
				|F_w|\le f}}}S_{F_u,F_w}$  from redundant paths in $G_{\overline{F_u}}$  in order to satisfy its \texttt{Maximal-Consistency}  condition during the parallel execution for set $F_u$. Similarly with previous arguments, by Theorem~\ref{thm:propagate} and the strong connectivity of $S_{F_u, F_w}$, there exists a simple $(q,c)$-path $p_{q,c}$ in $G_{\overline{F_u}}$, which propagates this  value $x_q$ to $c$.
	Since, by assumption, $u$  executes Filter-and-Average during its parallel execution for $F_u$, by the \texttt{Maximal-Consistency}  condition, $u$ will also receive the same  unique value $ x_q$, for each such node $q$,  propagated by redundant path $p_{q,c}|| p_{c,u}$ because $p_{c,u}$ is fully nonfaulty and entirely contained in $G_{\overline{F_u}}$. As argued previously, $v$ will receive $(\calm^c_v,COMPLETE(F_v))$ by a fully nonfaulty path $p_{c,v}$. Consequently, by Lemma~\ref{lemma: RCmain}, $v$ will satisfy the $Completeness(\calm_v,\calm^c, F_u)$  condition and thus, due to Definition~\ref{definition:informed}, $v$ will be informed about $F_u$. For the $Completeness(\calm_v,\calm^c, F_u)$  condition to be satisfied at node $v$, it must receive  the respective $x_q$ values which are 
	consistent with the ones in $\calm^c$, received through the fully nonfaulty path $p_{c,v}$. Thus, by definition of the $Completeness(\calm_v,\calm^c, F_u)$  condition, $v$ will also receive the same values $x_q$ for each $\displaystyle q\in \bigcup_{\mathclap{\substack{F_w\neq F_u\\
				|F_w|\le f}}}S_{F_u,F_w}$.

\end{proof}


%
%
%
%

Finally, we introduce notions that will be useful for our analysis later.

\begin{definition}
\label{def:common-info-set}
	Assume nonfaulty nodes $v,u$ which  execute Filter-and-Average during their parallel executions for sets $F_v$ and $F_u$, respectively.  Let $F_t\in \{F_v,F_u\}$ be the set about which both $v,u$ are informed and both receive a common value $ x_q$ for each $\displaystyle q\in \bigcup_{\mathclap{\substack{F_w\neq F_t\\
				|F_w|\le f}}}S_{F_t,F_w}$, as implied by Theorem~\ref{thm:valuesubset}. 
	We will refer to set $F_t$ as the \underline{common fault set} of $v,u$, and to $t$ as the \underline{leading node} of the pair. Considering the common values $x_q$ received by both $v,u$,
	for any set  $F_w\neq F_t$ with $F_w\subseteq V$, $|F_w|\le f$,  we define the \underline{common value set} $R_{F_w}$ as:
	\begin{equation}
R_{F_w}=\bigcup_{q\in S_{F_t,F_w}} x_q
	\end{equation}
	%
	
\end{definition}

%
%

\subsection{Value Update}

We next present Algorithm  \emph{Filter-and-Average} (FA), proposing a way for a node to filter its received messages in order to update its state value by an averaging procedure. Following the intuition of~\cite{SV17}, a node first sorts all the values in received message set $\calm_v$, which results to a sorted vector $O_v[r]$ in round $r$. In the next step, the node computes the maximal set of lowest values that might have been tampered by a faulty set (i.e., such that their propagation paths have an $f$-cover) and trims (removes) them from sorted vector $O_v[r]$. Analogously, the node also trims from $O_v[r]$ the maximal set of the highest values that may have been tampered by a faulty set. Finally, 
  Finally, the remaining sorted values, denoted as $O_v'[r]$, are averaged as is usual in the majority of the approximate consensus literature (e.g.,~\cite{AA_Dolev_1986,KA94, BPPT16}).


\begin{algorithm}
	\caption{\texttt{Filter-and-Average$(\calm_v)$} (for node $v$ in round $r$)}\label{alg:filteraverage}\medskip
	
	\KwIn{Incoming message history $\calm_v$ at the point where Filter-and-Average is called in BW in round $r$
	} \bigskip
	
	\nonl \hspace{-15pt}\textbf{ Code for $v \in V$:} \smallskip
	
	Sort all messages $m\in\calm_v$ in increasing order with respect to $value(m)$ which results in sorted vector $O_v[r]$.  
	
	$O_v^{lo}\leftarrow$ the longest message prefix of $O_v[r]$ for which $\exists$ an $f$-cover $F_{lo}$ of $\calp(O_v^{lo})$.
	
	$O_v^{hi}\leftarrow$ the longest message suffix of $O_v[r]$ for which $\exists$ an $f$-cover $F_{hi}$ of $\calp(O_v^{hi})$.\medskip

	\Comment*[f]{Trim extreme values and average}\smallskip
	
	Remove message 	$O_v^{lo},O_v^{hi}$ from $O_v[r]$ which results in the trimmed vector $O_v'[r]$.
	
	$x_v[r+1]=\frac{\max(O_v'[r])-\min(O_v'[r])}{2}$ \label{line:updatevalue}

\end{algorithm}	
~


Towards proving the convergence property, we will first show that for any nonfaulty nodes $v,u$ running Algorithm Filter-and-Average in round $r$, there will be a common value in the trimmed vectors $O'_v[r]$ and $O'_u[r]$ of the two nodes. For simplicity of presentation, we will omit the round variable $r$ in the following discussion.  The theorems presented next guarantee the existence of this common element. 

\subsection{Properties of Algorithm Filter-and Average}

According to the notation of Definition~\ref{def:common-info-set}, Theorem~\ref{thm:valuesubset} implies that for any pair of nonfaulty nodes $v,u$ executing Algorithm Filter-and-Average, there exists the common fault set  $F_t$ such that both of nodes $v$ and $u$ will be \textit{informed} about $F_t$ and will obtain the same common value sets $R_{F_w}$  for all $F_w\neq F_t$. 
The following theorem guarantees that for $F_t$,  there will be some source components whose values will appear in the trimmed vector of $v,u$ regardless of the sets $F_{lo}, F_{hi}$ used to trim their vectors. Recall that the notion of common fault set and leading node are introduced in Definition \ref{def:common-info-set}.

\begin{theorem}\label{theorem:aftertrim}
	Let $v,u$ be any pair of nonfaulty nodes, $F_t$ their common fault set, and $t\in \{v,u\}$ the leading node. Then for $z\in \{v,u\}$ and $f$-covers $F_{lo}^z, F_{hi}^z$ as identified in Algorithm~\ref{alg:filteraverage}, common value set $R_{F_{lo}^z}$ will be included in the vector $O_z$ after removing  $O_z^{lo}$,   and common value set $R_{F_{hi}^z}$ will be included in the vector $O_z$ after removing  $O_z^{hi}$.
\end{theorem}

\begin{proof}

	Without loss of generality, assume that $t=u$. We first consider the validity of the theorem for node $u$. The fact that $t=u$ implies that  $F_t=F_u$ is the set corresponding to the parallel execution during which $u$  executes Filter-and-Average. By Theorem~\ref{thm:propagate} and the strong connectivity of $S_{F_u, F_{lo}^u}$, if $u\notin S_{F_u, F_{lo}^u} $, it must have  received each value of $R_{F_{lo}^u}$ (as defined in Definition~\ref{def:common-info-set}) from $f+1$ disjoint $(S_{F_u, F_{lo}^u}, u)$-paths in $G_{\overline{F_u}}$ to satisfy its \texttt{Maximal-Consistency}  condition. Since $|F_{lo}^u|\le f$ there will be at least one path propagating each value of $R_{F_{lo}^u}$ in  $G_{\overline{F_u\cup F_{lo}^u}}$.  If $u\in S_{F_u, F_{lo}^u}$, by the strong connectivity of $S_{F_u, F_{lo}^u}$, $u$ will receive $R_{F_{lo}^u}$ from paths entirely within $S_{F_u, F_{lo}^u}$, i.e., paths in  $G_{\overline{F_u\cup F_{lo}^u}}$.  Thus, in both cases, common value set $R_{F_{lo}^u}$ will be included in vector $O_u$ after removing $O_u^{lo}$. Similar arguments hold for the case of $O_u^{hi}$.
	
	
	Next, we consider node $v$ and assume that it  executes Filter-and-Average during its parallel execution for $F_v$. Since $v\neq t=u$, node  $v$ has satisfied the  $Completeness(\calm_v,\calm^c, F_u)$   condition after receiving message $(\calm^c, COMPLETE(F_u))$ from a nonfaulty path initiating at $c\in reach_v(F\cup F_v)	\cap reach_u(F\cup F_u)$; this holds by an argument identical to that of the proof of Theorem~\ref{thm:valuesubset}. Similarly with the proof of Theorem~\ref{thm:valuesubset}, $c$ will propagate all values of $R_{F_{lo}^v}$ to $v$ through its FIFO-flooded message $(\calm^c, COMPLETE(F_u))$. Since $v$ satisfies $Completeness(\calm_v,\calm^c, F_u)$,  it receives each value of $R_{F_{lo}^v}$ through a path set $P$ \underline{with no $f$-cover} $H\subseteq V\setminus S_{F_u,F_{lo}^v}$. Consequently, since $|F_{lo}^v|\le f$ and $F_{lo}^v\cap S_{F_u,F_{lo}^v}=\emptyset$, one of the paths of $P$ will not contain any node in $F_{lo}^v$. This means that for any  value $x_q\in R_{F_{lo}^v}$, there exists a path in $G_{\overline{F_{lo}^v}}$ from which $v$ will receive $x_q$, and thus, all values of $R_{F_{lo}^v}$ will be included in $O_v^{lo}$.  Similar arguments hold for the case of $O_v^{hi}$.
\end{proof}

%

 
The next theorem facilitates the analysis later; it states that there is always an overlap between certain pairs of source components of reduced graphs. Its proof is deferred to the Appendix~\ref{sec:proof of overlap}.

\begin{theorem}\label{theorem:overlapsource}
	Suppose that graph $G=(V, E)$ satisfies condition 3-reach. For any three sets $F_v, F_u, F_w$, with  $F_v\subset V, F_u, F_w\subseteq V\setminus F_v$ and $|F_v|, |F_u|,|F_w|\le f$, $S_{F_v,F_u}\cap S_{F_v,F_w}\neq \emptyset$.
\end{theorem}

We next define some notions, helpful to determine the existence of  a common value in the intersection of $O'_v, O'_u$ for any pair of nonfaulty nodes $v,u$. As before, we assume that $F_t$  is the common fault set of $v,u$.


\begin{definition}\label{definition:minmaxvalues}
	
	Let $v,u$ be two nonfaulty nodes and $F_t$  their common fault set. For $F_w\subseteq V$ and $|F|\le f$, 	 let $\displaystyle x^{F_w}_{\min}=\min_{x\in R_{F_w}}x$, i.e.,  the minimum common value for the source component   $S_{F_t,F_w}$.
	Define the maximum of all these minimum values over all possible $F_w$ as
	$$x_{\max\min}=\max_{\mathclap{\substack{F_w\subseteq V\\
				|F_w|\le f}}} \  x_{\min}^{F_w}$$

	\noindent and let $S_{F_t,F_l}$, be a source component that includes the common value $x_{\max\min}$, i.e., $x_{\max\min}\in R_{F_l}$. 
	%
	%
	%
	%
	Analogously, let $x^{F_w}_{\max}$ be the maximum common value for the source component   $S_{F_t,F_w}$ and define minimum of these maximum values as
	$$x_{\min\max}=\min_{\mathclap{\substack{F_w\subseteq V\\
				|F_w|\le f}}} \  x_{\max}^{F_w}$$
	Similar to before, we assume that $S_{F_t,F_h}$ is a source component that includes the common value $x_{\min\max}$.
	
	%
	%
\end{definition}

%
%
%
%
%

\begin{lemma}\label{cor:difference of maxmin}
	For any two nonfaulty nodes $v,u$,   $x_{\max\min}\leq x_{\min\max}$
\end{lemma}

\begin{proof}
	By way of contradiction, assume that
	
	\begin{equation}
	    \label{eq:1}
	    x_{\max\min}> x_{\min\max}
	\end{equation}
	
	Then by Definition~\ref{definition:minmaxvalues}, we have the following two inequalities:
	
	\begin{equation}
	    \label{eq:2}
	    \text{For all values}~x~\text{ contained in}~R_{F_l},~~~x\ge x_{\max\min}
	\end{equation}
	
    \begin{equation}
	    \label{eq:3}
	    \text{For all values}~y~\text{ contained in}~R_{F_h},~~~y\ge x_{\min\max}
	\end{equation}
	
	Now we make two observations:
	
	\begin{itemize}
	    \item \textit{Observation 1}: Equations \ref{eq:1}, \ref{eq:2}, and \ref{eq:3} imply that $R_{F_l}\cap R_{F_h}=\emptyset$. 
	    
	    \item \textit{Observation 2}: By Definition \ref{def:common-info-set}, for each $w\in S_{F_t, F_l}\cap S_{F_t, F_h}$ there will be a common value $x_w$ contained in both $R_{F_l}$ and $R_{F_h}$.
	\end{itemize}
	These two observations imply that $S_{F_t, F_l}\cap S_{F_t, F_h}=\emptyset$, a contradiction to Theorem~\ref{theorem:overlapsource}.
\end{proof}

\begin{theorem}
\label{theorem:overlap}
	For nonfaulty nodes $v,u$, after the termination of Algorithm Filter-and-Average, we have  $O'_v\cap O'_u\neq \emptyset$.

\end{theorem}

\begin{proof}
	
	We will argue that a common value $x$ contained in $R_{F_l}\cap R_{F_h}$ will be included in both the trimmed vectors $O_z'$, for $z\in \{v,u\}$. 
	For any $F_{lo}^z$  with $|F_{lo}^z|\le f$ chosen by any $z\in \{v,u\}$, 
	define $x^{z, lo}_{\min}$ as the minimum value  contained in $R_{F_{lo}^z}$. Then by the definition of $x_{\max\min}$, we have $x^{z, lo}_{\min}\le x_{\max\min}$. 
	
	Due to Theorem~\ref{theorem:aftertrim}, value $x^{z, lo}_{\min}$ will be contained in $O_z$ after removal of set $O_z^{lo}$. Thus, due to the definition of $O_z^{lo}$, any value $x$ contained in $O_z^{lo}$ will satisfy $x\le x^{z, lo}_{\min}$, i.e., only values less or equal to $x^{z, lo}_{\min}$ will be removed from $O_z$ due to removal of $O_z^{lo}$ and the value $x^{z, lo}_{\min}$ will remain in the trimmed $O_z$. Note that, in the event that there are multiple values identical to $x^{z, lo}_{\min}$, then at least one instance of $x^{z, lo}_{\min}$ remains in $O_z$. 
	
	Next,  observe that for $z\in \{v,u\}$ and any choice of $F_{hi}^z$, $x^{z,hi}_{\max}\ge x_{\min\max}$ holds, where $x^{z,hi}_{\max}$ is the maximum value  contained  in $R_{F_{lo}^z}$, due to the definition of $x_{\min\max}$. Due to Theorem~\ref{theorem:aftertrim}, $x^{z,hi}_{\max}$ will be contained in $O_z$ after removal of set $O_z^{hi}$. Similarly with the previous argument, any value $x$ contained in $O_z^{hi}$ will satisfy $x\ge x^{z,hi}_{\max}$ and the value $x^{z, hi}_{\max}$ will remain in the trimmed $O_z$. Note that, in the event that there are multiple values identical to $x^{z, hi}_{\max}$, then at least one instance of $x^{z, hi}_{\max}$ remains in $O_z$.  Now, by Lemma~\ref{cor:difference of maxmin}  we have that,
	\begin{equation}\label{eq:inequality}
x^{z,lo}_{\min}\le x_{\max\min}\le x_{\min\max}\le x^{z,hi}_{\max}
	\end{equation}
	%
	Consequently, the only values removed from $O_z$ will be less or equal to $x^{z,lo}_{\min}$, greater or equal to $x^{z,hi}_{\max}$, and values $x^{z,lo}_{\min}, x^{z,hi}_{\max}$   will \textit{not} be trimmed. Thus, due to Equation (\ref{eq:inequality}), $x_{\max\min}$ and  $x_{\min\max}$ will be included in the final trimmed vector $O'_z$ for $z\in \{v,u\}$.
	
\end{proof}


\subsection{Correctness of Algorithm BW}
\label{s:correctness_proof}

For a given execution round $r\ge 0$, recall that $x_v[r]$ is the state variable maintained at node $v$ at the end of round $r-1$. Value $x_v[0]$ is assumed to be the input given to node $v$. 
We denote by $\M[r], \m[r]$,   the maximum and the minimum state value at nonfaulty nodes
by the end of round $r$. Since the initial state of each node is equal to its input, $\M[0]$ and $\m[0]$ is equal to the maximum and minimum value of the initial input of the nodes, respectively. Consequently, the convergence and validity requirements of approximate consensus can be stated as follows. 

\begin{itemize}
	\item \textit{Convergence }: $\forall \epsilon>0, \text{ there exists an iteration } r_\epsilon \text{ such that  for all } r\ge r_\epsilon, \M[r] - \m[r] <\epsilon$
	
	\item \textit{Validity}:   $\forall r>0,  \M[r] \le \M[0] \text { and } \m[r] \ge \m[0]$.
\end{itemize}

We next prove the convergence of the proposed algorithm which is based on the following lemma.

\begin{lemma}\label{lemma:conv}
	For every round $r$, it holds that
	\begin{align*}
		\frac{\M[r]-\m[r]}{2} \geq
		\M[r+1]-\m[r+1]
	\end{align*}
\end{lemma}

\begin{proof}
	For any $r > 0$, consider any pair of nonfaulty nodes $v, u$. 
	Without loss of generality, assume $x_v[r+1] \geq x_u[r+1]$. We prove the lemma by showing that 
	
	\begin{equation}
	    \label{eq:bound}
	    \frac{\M[r]-\m[r]}{2} \geq x_v[r+1] - x_u[r+1].
	\end{equation}
	
	%
	%
	%

	
	Observe that by Theorem~\ref{theorem:overlap}, the existence of element $z$ with $z\in O'_v[r] \cap O'_u[r]$ is proved. This implies that $\max(O'_u[r]) \geq z$.\footnote{Since $O'_u[r]$ is a sorted message set with respect to values, we define $\max(O'_u[r])$ and $\min(O'_u[r])$ to be  the maximum and minimum value respectively, included in this set as the first component of its messages-pairs.} Moreover,  $\min(O'_u[r]) \geq \m[r]$ holds, since if for node $u$, $F_{lo}=F$, then all actual faulty values are removed from from $O_u[r]$ resulting to trimmed vector $O'_u[r]$. Thus, for any remaining value $x$ in $O'_u[r]$, it holds  that $x\ge \m[r]$. If $F_{lo}\neq F$, then there exists a nonfaulty node $w$ such that $x_w[r]$, state value at node $w$ in round $r$, is trimmed, and hence, is smaller than any $x\in O'_u[r]$. 
	Therefore for any  $x\in O'_u[r]$, $x\ge \m[r]$ holds. Consequently, since $\max(O'_u[r]) \geq z$ and $\min(O'_u[r]) \geq \m[r]$,  due to line~\ref{line:updatevalue} of Algorithm Filter-and-Average, we have that

	\[
	x_u[r+1]=\frac{\max(O_v'[r])-\min(O_v'[r])}{2} \geq \frac{z+\m[r]}{2}
	\]
	Similarly, $\min(O'_v[r]) \leq z$ and $\max(O'_v[r]) \leq \M[r]$, which implies
	\[
	x_v[r+1] \leq \frac{z+\M[r]}{2}
	\] 
	Equation (\ref{eq:bound}) follows from these two inequalities.
\end{proof}

\noindent Lemma~\ref{lemma:conv} and simple arithmetic operations 
 imply the Convergence property (details appear in the termination study of Algorithm BW presented below).
Validity is based on the observation that all the extreme values will be eliminated by each nonfaulty node owing to the way the trimmed vector $O_v'[p]$ is derived. The arguments are similar to that of the proof of Lemma~\ref{lemma:conv}. The proof is presented Appendix~\ref{sec:validity}.


\begin{theorem}[Validity]\label{theorem:validity}
	$\forall r \ge 0, \M[r]\le  \M[0]$ and $\m[r] \ge \m[0]$
\end{theorem}

\commentall{
\begin{proof}
	We prove the claim by induction on the round index $r$. For $r=0$ both inequalities trivially hold.  Assume that the claim holds for round $r$. For any nonfaulty node $u$ we can show that   $\min(O'_u[r]) \geq \m[r]$ and $\max(O'_u[r]) \leq \M[r]$ with identical arguments used in the proof of Lemma~\ref{lemma:conv}. Consequently, we have that since $x_u[r+1]=\frac{\max(O_v'[r])-\min(O_v'[r])}{2} $, $ \m[r] \le x_u[r+1]\le \M[r]$ holds.
\end{proof}
}


\subsubsection*{Termination of Algorithm BW}
Recall that the termination requirement for approximate consensus (Definition~\ref{definition:approx}) requires that all nonfaulty nodes should output a value. 
We follow the approach in \cite{Tseng_podc2015,TV12arxiv}.
Suppose that the input is within the range $[0, K]$, where $K \in \mathbb{R}$ is known \textsl{a priori}.
If $K <\epsilon$ , then the problem is trivial, so it is assumed that  $K\ge \epsilon$. Repeated application of Lemma~\ref{lemma:conv} implies that $\M[r+1]-\m[r+1]\le \frac{\M[0]-\m[0]}{2^{r}}\le \frac{K}{2^{r}} $. This implies that for given $K, \epsilon$, the state values of the nonfaulty nodes will be within $\epsilon$ of each other after round $r>\log_2 \frac{K}{\epsilon}$. Since $K,\epsilon$ are \textsl{a priori known }, each node can locally compute $\frac{K}{\epsilon}$ and output its value in the first round $r$ such that $r>\log_2 \frac{K}{\epsilon}$.
 


\begin{acks}
Nitin Vaidya's work is supported in part by the Army Research Laboratory under Cooperative Agreement W911NF-17-2-0196, and by the National Science Foundation award 1842198. The views and conclusions contained in this document are those of the authors and should not be interpreted as representing the official policies, either expressed or implied, of the Army Research Laboratory, National Science Foundation or the U.S. Government.
\end{acks}


\bibliographystyle{ACM-Reference-Format}
\bibliography{bibliography}

\newpage
\appendix

\section{The $k$-reach condition family}
\label{sec:kreach}
In this section, we summarize the tight topological conditions related with consensus in directed networks that have appeared in the literature along with their equivalents from the family of $k$-reach conditions, for $k=1,2,3$. The topological conditions {\bf CCS} (abbreviating Crash-Consensus-Synchronous), {\bf CCA}
(Crash-Consensus-Asynchronous) and {\bf BCS} (Byzantine-Consensus-Synchronous)  were introduced in~\cite{Tseng_podc2015} and proven tight for the cases of synchronous  crash  consensus, asynchronous approximate crash  consensus and synchronous Byzantine  consensus respectively. The determination of the necessary and sufficient topological condition for solving approximate Byzantine consensus in asynchronous systems, has been an open problem since {2012}.

We next present some additional  definitions that facilitate our presentation.

 For set $B\subseteq V$, process $v$ is said to be an \emph{incoming  (resp. outgoing)
	neighbor of set $B$}  if $v\not\in B$, and there exists $u\in B$
such that $(v,u)\in E$ (resp. $(u,v)\in E$). The \emph{incoming} and \emph{outgoing neighborhood of a node} $v$  are the sets of its incoming and outgoing neighbors respectively and will be denoted with $\caln^-_v, \caln^+_v$ respectively. We extend the notion to the \emph{incoming (resp. outgoing) neighborhood of a set} $B$, denoted with $\caln^-_B$ (resp. $\caln^+_B$) and defined as the set of all incoming (resp. outgoing) neighbors of $B$.   Given subsets of nodes $A$ and $B$, set $B$ is said to \emph{ have $k$ incoming neighbors in set $A$ } if $A$ contains $k$ distinct incoming neighbors of $B$. Next, we define a notion which concerns the connectivity of any two node sets of the graph as presented in~\cite{Tseng_podc2015}.

\begin{definition}
	\label{def:point}
	Given disjoint non-empty subsets of nodes $A$ and $B$, we will say that $\point{A}{B}{x}$ holds, if $B$ has at least $x$  incoming neighbors in $A$. The negation of $\point{A}{B}{x}$ will be denoted by $\notpoint{A}{B}{x}$. 
\end{definition}

We also introduce the following useful generalization fo the reach set notion. The notion denotes all the multi-hop incoming neighbors of node $v$ in graph $G_{\fbar}$.

\begin{definition}[Reach set of $v$ under $F$]\label{def:reachappendix}
	For a subgraph $G'=(V',E')$ of $G$, node $v\in V'$ and  node set $F\subseteq V \setminus\{v\}$, we will use the following notation,
	$$reach_v^{G'}(F)=\{u\in V'\setminus F : \text{ $v$ is reachable from $u$ in graph } G'_{V'\setminus F}  \}$$
	
	Whenever $G'=G$ we will omit the superscript $G'$ and simply use the notation $reach_v(F)$. 
	
\end{definition}

The definitions of conditions CCS, CCA, BCS defined in~\cite{Tseng_podc2015} follow.

\begin{definition}[{\bf Condition CCS}]
	\label{t_ccs1}
	For any partition $F, L, C, R$ of $V$, where $L$ and $R$ are both non-empty, and $|F| \leq f$, at least one of the following  holds:
	\begin{itemize}
		\item  $\point{L \cup C}{R}{1}$
		\item $\point{R \cup C}{L}{1}$
	\end{itemize} 
\end{definition}

\begin{definition}[{\bf Condition CCA}]
	\label{t_cca1}
	For any partition $L, C, R$ of $V$, where $L$ and $R$ are both non-empty, at least one of the following  holds:
	\begin{itemize}
		\item  $\point{L \cup C}{R}{f+1}$
		\item $\point{R \cup C}{L}{f+1}$
	\end{itemize} 
\end{definition}

\begin{definition}[{\bf Condition BCS}]
	\label{t_bcs1}
	For any partition $F, L, C, R$ of $V$, where $L$ and $R$ are both non-empty, and $|F| \leq f$, at least one of the following  holds:
	\begin{itemize}
		\item  $\point{L \cup C}{R}{f+1}$
		\item $\point{R \cup C}{L}{f+1}$
	\end{itemize} 
\end{definition}

Observe that while BCS requires a 4-set partition $F,L,R,C$ of $V$, condition CCA only requires a 3-set partition $L,R,C$ of $V$.

We next present an equivalent condition to CCS, CCA, BCS,  based on reach sets of any two nodes of the graph. 

\begin{definitions}\label{def:reach2}
	In the following, sets $F,F_v, F_u\subseteq V$ intuitively represent possible faulty sets and thus are of cardinality at most $f$, i.e., $|F|, |F_v|, |F_u|\le f$. We define the three following conditions,
	
	\begin{itemize}
		\item \textbf{1-reach:} For any $F\subset V$ such that $|F|\leq f$ and any nodes $u,v\in \overline{F}$, we have $$reach_u(F)\cap reach_v(F)\neq \emptyset.$$
		
		\item \textbf{2-reach:} For any nodes $u, v\in V$ and any node subsets $F_u$, $F_v$ such that $|F_u|, |F_v|\leq f$, $u\in \overline{F_u}$, and $v \in \overline{F_v}$, we have $$reach_v(F_v)\cap reach_u(F_u)\neq \emptyset.$$
		
		\item \textbf{3-reach:}
		For any nodes $u, v\in V$ and any node subsets $F$, $F_u$, $F_v$ such that $|F|, |F_u|, |F_v|\leq f$, $u\in \overline{F\cup F_u}$, and $v \in \overline{F\cup F_v}$, we have
		$$reach_v(F\cup F_v)\cap reach_u(F\cup F_u)\neq \emptyset.$$
	\end{itemize}

\end{definitions}

\noindent Observe that in a clique, it holds that $reach_v(F_v) \cap  reach_u(F_u) = reach_v(F_v\cup~F_u)\cap reach_u(F_v\cup F_u)$. Thus, for example condition 3-reach in a clique is equivalent with,
$$reach_v(F\cup F_v\cup F_u)\cap reach_u(F\cup F_v\cup F_u)\neq \emptyset$$ which is equivalent with the well known clique condition $n> 3f$, tight for byzantine consensus. Analogously, one can show that in a clique, 1-reach and 2-reach are equivalent with $n>f$ and $n>2f$ respectively.

We next present the generalization of the above conditions $k$-reach which determines the family of conditions encompassing the above. 

\begin{definition}
For any sets $F, F_v^1,F_u^1,  \ldots, F_v^k, F_u^k$, each of cardinality at most $f$

$$\textbf{\textbf{k-reach:} }\begin{cases}
	reach_v(F_v^1\cup\ldots \cup F_v^k)\cap reach_u(F_u^1\cup\ldots \cup F_u^k)\neq \emptyset & \text{ if } k= even\\
	reach_v(F\cup F_v^1\cup\ldots \cup F_v^{k-1})\cap reach_u(F\cup F_u^1\cup\ldots \cup F_u^{k-1})       & \text{ if } k= odd
\end{cases}$$
	
\end{definition}

In the following theorem, we show that conditions 1-reach, 2-reach, and 3-reach prove are equivalent to CCS, CCA, and BCS respectively.

\begin{theorem} \label{thm:reachequivalence}
	\leavevmode
	\begin{enumerate}[label=(\alph*) ]
		\item CCS$\Leftrightarrow$ 1-reach
		\item CCA$\Leftrightarrow$ 2-reach
		\item BCS$\Leftrightarrow$ 3-reach
	\end{enumerate}
	
\end{theorem}

\begin{proof}
	\leavevmode
	
	$(a)$ Condition 1-reach is trivially equivalent with the existence of a directed rooted tree in $G_{\fbar}$ as presented in~\cite{Tseng_podc2015}. In turn, the equivalence of the latter condition with CCS has been proven in~\cite{Tseng_podc2015, ST19book}.\medskip
	
	$(b)$ Direction``$\Rightarrow$'' is implicitly proven in~\cite{Tseng_podc2015}\footnote{The claim is implied by the proof of Lemma 7 in~\cite{Tseng_podc2015}} proves the $``\Rightarrow"$ direction. We next prove direction ``$\Leftarrow$``. \\
	If CCA does not hold in $G$, then there exists a partition $L,R,C$ of $V$ with $L,R\neq \emptyset$ such that $\notpoint{L \cup C}{R}{f+1}$   and $\notpoint{R \cup C}{L}{f+1}$.
	%
	Observe that $|N^-_L|, |\caln^-_R|\le f$. This is because $L,R,C$ is a partition of $V$ and thus, $N^-_L\subseteq R\cup C$ and $N^-_R\subseteq L\cup C$; since CCA is not satisfied, the claim holds. Subsequently, let $v\in L, u\in R$; these nodes exist since $L,R\neq \emptyset$ as per CCA definition.  Note that there exist two sets $F_v=\caln^-_L, F_u=\caln^-_R$ of cardinality at most $f$ such that the following holds,
	$$reach_v(\caln^-_L)	 \cap reach_u(\caln^-_R)\subseteq L\cap R=\emptyset$$
	and thus condition 2-reach does not hold.\medskip
	
	$(c)$  For any sets $F, F_v\subseteq V$ with $|F|, |F_v|\le f$  and node $w \in V\setminus F\cup F_v$ we will use  $reach_w^{G_{\fbar}}(F_v)$ as defined in Definition~\ref{def:reach1}.  Note that by definitions~\ref{t_cca1} and~\ref{t_bcs1}, Condition BCS is equivalent to the following condition: for all sets $F\subseteq V$ with $|F|\le f$, CCA holds in $G_{\fbar}$.
	
	Due to the equivalence of CCA with 2-reach in the previous step $(b)$, it holds that BCS is equivalent with the following condition: For all sets $F, F_v, F_u\subseteq V$ with $|F|, |F_v|, |F_u|\le f$, 
	
	$$reach_v^{G_{\fbar}}(F_v)\cap reach_v^{G_{\fbar}}(F_v)\neq \emptyset$$
	Thus BCS is equivalent to,
	$$reach_v(F\cup F_v)\cap reach_v(F\cup F_v)\neq \emptyset $$
	which coincides with condition 3-reach.
	
\end{proof} 

\section{Necessity of condition 3-reach 
}
\label{sec:necessity}
We next show that condition 3-reach is necessary for asynchronous byzantine approximate consensus to be achieved in a network. With $e\stackrel{v}{\sim} e'$ we will denote the fact that  that execution $e$ is \emph{indistinguishable} from execution $e'$ with respect to node $v$ (cf.~\cite{AA_nancy}).
Note that, considering an approximate consensus algorithm,  $e\stackrel{v}{\sim} e'$ implies that node $v$ will output the same value in both executions $e, e'$.
To facilitate the proof, for $A,B\subseteq V$ we will use the notation $E(A,B)=\{(v,u): v\in A, u \in B\}$ to denote all  edges from set $A$ to set $B$.

\begin{theorem}[Impossibility of Approximate Consensus]
	Byzantine asynchronous approximate consensus is impossible in networks where condition 3-reach is not satisfied.
\end{theorem}

\begin{proof}
Consider a network $G=(V,E)$ where condition 3-reach is not satisfied and assume the existence of algorithm $\cala$ that achieves asynchronous approximate consensus in $G$. This means that there exist sets $F, F_v, F_u\subseteq V$ with $|F|, |F_v|, |F_u|\le f$, and  nodes $v\in \overline{F\cup F_v}$, $u\in \overline{F\cup F_u}$ such that:
\begin{equation}\label{eq:not3reach}
reach_v(F\cup F_v)\cap reach_u(F\cup F_u)=\emptyset
\end{equation}

	We  define the following three executions of $\cala$ determined by the set of faulty nodes and their behavior, the inputs of all nodes and the communication delay. Observe that a Byzantine fault may simply deviate from the protocol by crashing and not sending any message since the Byzantine fault model is strictly stronger than the crash fault. Also, note that we can assume an external notion of time (global clock), not directly observable by the nodes, to facilitate the explicit description of delays. The latter technique has been considered in~\cite{GGGK13, ST18}.
	
	\begin{enumerate}[label={($e_{\arabic*}$)} ]
		\item The input of every node $z\in V$ is $x_z=0$, all nodes in $F_v$ have crashed from the beginning of the execution and all other nodes are nonfaulty; the latter is possible since $|F_v|\le f$.

		\item The input of every node $z\in V$ is $x_z=\epsilon$, all nodes in $F_u$ have crashed from the beginning of the execution and all other nodes are nonfaulty; the latter is possible since $|F_u|\le f$.  
		
		\item \emph{Inputs:} The input of every node $z\in reach_v(F\cup F_v)$ is $x_z=0$, the input of every node $w\in reach_u(F\cup F_u)$ is $x_w=\epsilon$; these inputs are well defined because of Eq.~\ref{eq:not3reach}. All remaining nodes in $V\setminus (reach_v(F\cup F_v)\cup reach_u(F\cup F_u))$ have arbitrary inputs.\smallskip
		
		\emph{Delivery delays:}   Message deliveries delays are the same as $e_1$ and $e_2$ except the delays of all messages transmitted through edges $E(F_v, reach_v(F\cup F_v))$, and all messages transmitted  through edges $E(F_u, reach_u(F\cup F_u))$. We assume that the delivery delay of the latter messages is lower bounded by an arbitrary number $T\in \mathbb{N}$ of time-steps. The exact value of $T$ will be defined in the following. Message deliveries though all other edges are instant.\smallskip 
		
		\emph{Faulty set behavior:} Node set $F$ is faulty and behaves towards $reach_v(F\cup F_v)$ as in $e_1$ and towards $reach_u(F\cup F_u)$ as in $e_2$. More concretely, all messages transmitted through edges in $E(F, reach_v(F\cup F_v) )$ are identical to the messages transmitted through $E(F, reach_v(F\cup F_v) )$ in $e_1$ and all messages transmitted through edges in $E(F, reach_u(F\cup F_u) )$ are identical to the messages transmitted in $e_2$. Observe that Eq~\ref{eq:not3reach} implies that $E(F, reach_v(F\cup F_v) )\cap E(F, reach_u(F\cup F_u) )=\emptyset$ and thus the latter behavior is well defined. This holds because if there exists an edge $(w,z) \in E(F, reach_v(F\cup F_v) )\cap E(F, reach_u(F\cup F_u) ) $ , then $w\in F$ and $z\in reach_v(F\cup F_v) \cap reach_u(F\cup F_u)$; the latter contradicts Eq,~\ref{eq:not3reach}.  Also, this  behavior is possible under the Byzantine faults model since $|F|\le f$ and Byzantine faults can have any arbitrary behavior~\footnote{This is a standard argument used towards indistinguishability in distributed systems. Details proving that this  faulty behavior  is well defined can be found in~\cite{PPS17}.}. 
	\end{enumerate}
	
	Note that all three executions are well defined due to the fact that 3-reach condition is not satisfied (Eq.~\ref{eq:not3reach}). Since $e_1$ is a well defined execution of $\cala$, there will be a specific time point $t_L$ by which, node $v$ will terminate in $e_1$. Moreover, in order to satisfy the validity condition $v$ will output the value $0$ upon termination of  $e_1$.  Similarly, since $e_2$ is a well defined execution of $\cala$, there will be a specific time point $t_R$ by which, node $u$ will terminate in $e_2$ by outputting value $\epsilon$. We now assume that the lower bound $T$ for all delivery delays described in execution $e_3$ is any $T$ with 
	$$T>\max\{t_L,t_R\}$$
	
	Now consider execution $e_3$;  all messages received by $v$ are the same in executions $e_1, e_3$. Therefore   $e_3\stackrel{v}{\sim} e_1$ holds and by the previous argument $v$ will output $0$ in execution $e_3$. Similarly  $e_3\stackrel{u}{\sim} e_2$ holds and $u$ will output $\epsilon$ in execution $e_3$. Thus, the convergence property is violated.
	
\end{proof}

\section{Proof of Theorem~\ref{theorem:overlapsource}}

\label{sec:proof of overlap}

The next theorem states that there is always an overlap between certain pairs of source components of reduced graphs. For ease of presentation we prove the theorem for condition BCS, which was proved equivalent to 3-reach in Theorem~\ref{thm:reachequivalence}.

\begin{theorem}
	Suppose that graph $G=(V, E)$ satisfies condition 3-reach. For any three sets $F_v, F_u, F_w$, with  $F_v\subset V, F_u, F_w\subseteq V\setminus F_v$ and $|F_v|, |F_u|,|F_w|\le f$, $S_{F_v,F_u}\cap S_{F_v,F_w}\neq \emptyset$
\end{theorem}

\begin{proof}
	
	Observe that by definition of a source component of a reduced graph it holds that 
	\begin{align}
	\caln^-_{S_{F_v,F_u}}\subseteq F_v\cup F_u \label{eq1: neighb}\\
	\caln^-_{S_{F_v,F_w}}\subseteq F_v\cup F_w
	\end{align}
	
	If $S_{F_v,F_u}\cap S_{F_v,F_w}=\emptyset$ then we can consider the following partition $L,R,F,C$ of $V$.
	
	\begin{align*}
	L&=S_{F_v,F_u} \\
	R&=S_{F_v,F_w}\\
	F&=F_v\\
	C&=V\setminus (L\cup R\cup F)	\end{align*}
	
	We will next show that $\notpoint{R \cup C}{L}{f+1}$ and $\notpoint{L \cup C}{R}{f+1}$. First, observe that $L,R\neq \emptyset$  by the definition of source component and the fact that BCS holds. Moreover, we have that,
	
	$$(R\cup C) \cap F=(S_{F_v,F_w}\cup (V\setminus S_{F_v,F_u} \setminus S_{F_v,F_w} \setminus F_v))\cap F_v=S_{F_v,F_w}\cap F_v=\emptyset$$
	where the latter equation holds by definition of $S_{F_v,F_w}$. Thus, by equation~\ref{eq1: neighb} we have that $N^-_L\cap (R\cup C)\subseteq F_u $, which is equivalent to $|N^-_L\cap (R\cup C)|\le f$, which in turn implies that $\notpoint{R \cup C}{L}{f+1}$. Similarly, $\notpoint{L \cup C}{R}{f+1}$ can be shown. These two facts imply that partition $L, R, F, C$ violates condition BCS; a contradiction since BCS is equivalent with 3-reach by Theorem~\ref{thm:reachequivalence}. 
\end{proof}

\section{Proof of Theorem~\ref{thm: FRtermination} }\label{sec:BWtermination}

\begin{theorem}
	Any nonfaulty node $v$ will eventually execute Filter-and-Average during a parallel execution for a set $F'$.
	
\end{theorem}

\begin{proof}
	Assume that $v$ does not  execute Filter-and-Average for any parallel execution. Due to lines~\ref{line:ifnextround}-\ref{line:nextroundtrue} of BW, this means that the shared variable $nextround$ will be false.  Consider the parallel execution for $F'=F$, where $F$ is the actual faulty set. 
	Next observe that due to Lemma~\ref{lemma: MCtermination}, $v$ will satisfy \texttt{Maximal-Consistency}  condition. The same holds for all other nonfaulty nodes.
	Consequently, since  all nodes in $reach_v(F)$ are nonfaulty, they will all eventually FIFO-Flood $COMPLETE(F)$ 
	and $v$ will eventually FIFO-receive all of these messages through nonfaulty paths in $G_{\overline{F}}$, satisfying the \texttt{FIFO-Receive-All} condition. Finally, by  Lemma~\ref{lemma: RCmain}, $v$ will eventually satisfy the $Completeness(\calm^c, F_u)$ condition corresponding to any received $(\calm^c, COMPLETE(F_u))$ message, since all these messages will be received through fully nonfaulty paths and thus function $Verify(\calm_v)$ will be assigned to true. Finally, since the $nextround$ variable  is assigned to false, node $v$ will  execute Filter-and-Average.  	
\end{proof}

\section{The Redundant Flood algorithm (\texttt{RedundantFlood)}}
\label{sec:rfloodalg}
	
	We next present the natural algorithm for flooding a message throughout all redundant paths in the network. To avoid a trivial adversary attack where the adversary lies about the propagation path, each time a node $v$ receives a message $(x,p)$ from node $u$ we assume that $v$ checks if $ter(p)=u$ and rejects the message if this is not the case. This is a feasible check since edges represent reliable communication channels where the recipient knows the identity of the sender. 

	
	
	~

	\begin{algorithm}[H]
		\caption{{RedundantFlood }} \label{alg:flood}\medskip
		\KwIn{sender node identifier $s$, sender's value $x$}\bigskip
		
		\textbf{Code for $s$:} send message $(x, \langle s \rangle)$ to all outgoing neighbors. 	\Comment*[f]{local broadcast} \medskip
		
		\textbf{Code for $v\neq s$:}\smallskip
		
		\If{$m=(x, p)$ is the first  message with $path(m)=p$ received from a node $u\in N^-_v$}{
			\If{$w\in \caln_v^+ \text{ and } p||v||w$ {is a redundant path}  }{
				send message $(x,p||v)$  to $w$
			}}\medskip
			
		\end{algorithm}

\section{The FIFO-Flood and FIFO-Receive procedures}
\label{sec:fifoflood}
In the following we present a high-level description of the \texttt{FIFO Flood} and \texttt{FIFO Receive}  procedures used in algorithm~\ref{alg:BW}. 

		\texttt{FIFO-Flood}. During this procedure, each node $i$ maintains a \emph{FIFO-counter} which is incremented any time the node sends a message. The counter is appended to every  message sent by node $i$.  We stress that this counter is a shared variable between all parallel threads at node $i$.
	
	\texttt{FIFO-Receive}. We will say that node $v$ FIFO-receives a message $m$  from  node $u$ propagated through a FIFO-Flood procedure if $v$ has also received all previous messages (with respect to FIFO-counter) sent by $u$. More concretely, if $v$ FIFO-receives $m$ with FIFO-counter $k$ initiated by node $u$, then $v$ must have received all messages initiated by $u$ with FIFO-counters $1,\ldots,k-1$.

	The latter procedures clearly implement FIFO channels between nonfaulty nodes which are connected via fully nonfaulty paths.  Trivially, the ordering of messages propagated through fully nonfaulty paths will be maintained since all FIFO-counters will be propagated correctly.  In the case of a path containing a faulty node, it is obvious that message order is impossible to maintain since the adversary can change the order arbitrarily under any protocol; this holds since the adversary can filter all information propagated through this path. 
	
	\section{Proof of Theorem~\ref{theorem:validity}}
	\label{sec:validity}
	
	\begin{theorem}[Validity]
	$\forall r \ge 0, \M[r]\le  \M[0]$ and $\m[r] \ge \m[0]$
\end{theorem}

\begin{proof}
	We prove the claim by induction on the round index $r$. For $r=0$ both inequalities trivially hold.  Assume that the claim holds for round $r$. For any nonfaulty node $u$ we can show that   $\min(O'_u[r]) \geq \m[r]$ and $\max(O'_u[r]) \leq \M[r]$ with identical arguments used in the proof of Lemma~\ref{lemma:conv}. Consequently, we have that since $x_u[r+1]=\frac{\max(O_v'[r])-\min(O_v'[r])}{2} $, $ \m[r] \le x_u[r+1]\le \M[r]$ holds.
\end{proof}

\end{document}